\documentclass[12pt]{article}
\usepackage[utf8]{inputenc}
\usepackage[british]{babel}
\usepackage{cmap}
\usepackage{lmodern}
\usepackage[T1]{fontenc}

\usepackage{amssymb, amsmath, amsthm}
\usepackage[a4paper,top=25mm,bottom=25mm,left=25mm,right=25mm]{geometry}
\usepackage{ragged2e}

\usepackage{authblk} 
\usepackage{pifont}
\usepackage{graphicx}
\usepackage[dvipsnames,svgnames,table]{xcolor}
\usepackage[figuresright]{rotating}
\usepackage{xtab} 
\usepackage{longtable} 
\usepackage{multirow}
\usepackage{footnote}
\usepackage[stable]{footmisc}
\usepackage{chngpage} 
\usepackage{pdflscape} 
\usepackage[nottoc,notlot,notlof]{tocbibind} 

\usepackage{pgfplots}
\pgfplotsset{every tick label/.append style={font=\footnotesize}}
\pgfplotsset{compat=1.18}
\usepackage{setspace}

\usepackage{array}
\newcolumntype{K}[1]{>{\centering\arraybackslash$}p{#1}<{$}}

\makesavenoteenv{tabular}
\usepackage{tabularx}
\usepackage{booktabs}
\usepackage{threeparttable}
\usepackage[referable]{threeparttablex} 
\newcolumntype{R}{>{\raggedleft\arraybackslash}X}
\newcolumntype{L}{>{\raggedright\arraybackslash}X}
\newcolumntype{C}{>{\centering\arraybackslash}X}
\newcolumntype{A}{>{\columncolor{gray!25}}C}
\newcolumntype{a}{>{\columncolor{gray!25}}c}

\newlength{\tablen}

\usepackage{dcolumn} 
\newcolumntype{.}{D{.}{.}{-1}}

\usepackage{tikz}
\usetikzlibrary{arrows, calc, matrix, patterns, positioning, trees}
\usepackage[semicolon]{natbib} 
\usepackage[hyphens]{xurl}
\usepackage[nopatch=footnote]{microtype}
\usepackage[justification=centering]{caption} 

\usepackage[labelformat=simple]{subcaption}

\DeclareCaptionLabelFormat{parenthesis}{(#2)}
\makeatletter
\renewcommand\p@subfigure{\arabic{figure}.}
\makeatother

\DeclareCaptionLabelFormat{parenthesis}{(#2)}
\makeatletter
\renewcommand\p@subtable{\arabic{table}.}
\makeatother

\usepackage{enumitem}

\setlist[itemize]{leftmargin=2.5\parindent}
\setlist[enumerate]{leftmargin=2.5\parindent}

\usepackage{hyperref} 
\hypersetup{
  colorlinks   = true,    		
  urlcolor     = blue,    		
  linkcolor    = blue,    		
  citecolor    = ForestGreen	
}

\def\addlegendimage{\csname pgfplots@addlegendimage\endcsname}

\theoremstyle{plain}

\theoremstyle{definition}

\theoremstyle{remark}

\makeatletter
\let\@fnsymbol\@alph
\makeatother

\def\keywords{\vspace{.5em} 
{\noindent \textit{Keywords}: }}

\def\AMS{\vspace{.5em} 
{\noindent \textbf{\emph{MSC} class}: }}

\def\JEL{\vspace{.5em} 
{\noindent \textbf{\emph{JEL} classification number}: }}

\title{Match classification in the last round \\ of four-team round-robin tournaments}

\author{\href{https://sites.google.com/view/laszlocsato}{L\'aszl\'o Csat\'o}\thanks{~Corresponding author. Email: \emph{laszlo.csato@sztaki.hun-ren.hu} \newline
Institute for Computer Science and Control (SZTAKI), Hungarian Research Network (HUN-REN), Laboratory on Engineering and Management Intelligence, Research Group of Operations Research and Decision Systems, Budapest, Hungary \newline
Corvinus University of Budapest (BCE), Institute of Operations and Decision Sciences, Department of Operations Research and Actuarial Sciences, Budapest, Hungary}
$\qquad \qquad$
Andr\'as Gyimesi\thanks{~Email: \emph{gyimesi.andras@ktk.pte.hu} \newline
University of P\'ecs, Faculty of Business and Economics, P\'ecs, Hungary \newline
Institute for Computer Science and Control (SZTAKI), Hungarian Research Network (HUN-REN), Laboratory on Engineering and Management Intelligence, Research Group of Operations Research and Decision Systems, Budapest, Hungary}}

\date{\today}

\def\Dedication{
{\noindent
``\emph{The FIFA Council has unanimously decided in favour of expanding the FIFA World Cup\textsuperscript{TM} to a 48-team competition as of the 2026 edition.
[\dots]
The decision was taken following a thorough analysis, based on a report that included four different format options. The study took into account such factors as sporting balance, competition quality, impact on football development, infrastructure, projections on financial position and the consequences for event delivery.}''
}
\vspace{0.25cm}

\flushright
\noindent
\small{(Third meeting of the FIFA Council \citep{FIFA2017d})}

\vspace{1cm} 
\justify }

\begin{document}
\newgeometry{top=20mm,bottom=25mm,left=25mm,right=25mm}
\maketitle

\thispagestyle{empty}
\Dedication

\begin{abstract}
\noindent
Classification of matches played in the last rounds of sports competitions is a well-established tool for evaluating tournament designs. Our paper offers the first comparison of deterministic and probabilistic match classification models by analysing the most prominent example of four-team round-robin competitions, the group stage of the FIFA World Cup. We show that both models are highly relevant in practice: all (four) deterministic and (six) probabilistic match types occurred in the 2014 and 2018 FIFA World Cups, respectively. The probabilistic model, which accounts for the relative benefits of attacking and defending, provides a deeper insight; for instance, the competitive matches from the deterministic approach can be of any of the six probabilistic types. Therefore, the probabilistic model is used to decompose the impact of three reforms introduced in the 2026 FIFA World Cup: the expansion to 48 teams, the modified qualification rules, and head-to-head tie-breakers.


\AMS{62P20, 90-10, 90B90, 91B14}

\JEL{C44, C53, D71, Z20}
\end{abstract}

\clearpage
\restoregeometry

\section{Introduction} \label{Sec1}

Various methods of operational research are increasingly used to examine sports rules. Indeed, as clearly shown by the justification of the F\'ed\'eration Internationale de Football Association (FIFA) for the expansion of the 2026 FIFA World Cup, quoted at the beginning of the paper, the evaluation of a tournament format requires an appropriate choice of the objective function and a well-designed quantification of several, often conflicting factors.
Inspired by the fundamental changes in the 2026 FIFA World Cup, the flagship tournament of the most popular sport in the world, our paper offers such an in-depth analysis.

In the history of football, there are several examples of tacit coordination \citep{Guyon2020a} and matches where one team or even both teams had few incentives to win. The last round of the group stage is particularly prone to these anomalies \citep{Guyon2022a, Csato2025b}.
Let us see some examples from the FIFA World Cup:
\begin{itemize}
\item 
In 2018, Denmark and France played a boring goalless draw, which secured the qualification of Denmark and the first position of France. The Danish coach said that ``\emph{we just needed one point, we were up against one of the best teams in the world at counterattacks, so we would have been stupid to open up a lot of space}'', and ``\emph{the last 15 minutes was a neutral match}'' according to the French manager \citep{Stuart2018}.
\item 
In 2018, Belgium and England played against each other for the first place after both teams had already qualified. England made eight changes from the winning squad against Panama, lost the game, and the coach admitted ``\emph{our objective was to qualify from the group and we have done that}'' \citep{BBC2018}.
\item
In 2022, Brazil qualified before the last round and completely shuffled the squad by 10 changes. They lost to Cameroon, and finished as the group winner only because Switzerland failed to score one more goal against Serbia in the parallel match \citep{BBC2022}.
\item
In 2026, France and Norway played against each other for the first place after both teams had already qualified. Norway opted to make 10 changes, the ``goal machine'' Haaland started on the bench for the first time since 2024, and the coach clarified that ``\emph{I could not care too much about that game now}'' \citep{Johnston2026}.
\end{itemize}
The academic literature has already devoted serious effort to classifying these matches according to their competitiveness. Inspired by the matches above, we adopt the common assumption of this research field that teams only focus on qualifying for the next stage, and do not care about the knockout bracket where they land.

However, there exist two different match classification approaches.
Deterministic models \citep{RibeiroUrrutia2005, ChaterArrondelGayantLaslier2021, CsatoMolontayPinter2024, DevriesereGoossensSpieksma2026} focus on mathematical (im)possibility and do not take the relative payoffs from winning and losing into account. A recently developed probabilistic model \citep{CsatoGyimesi2026b} is based on two premises:
(1) teams do not consider events with a small probability of occurring; and
(2) their offensive or defensive tactic is chosen based on the incentives provided by the tournament rules.

Our first contribution resides in the comparison of the deterministic and probabilistic match classification models. This is carried out for four-team single round-robin tournaments where the top two teams qualify, a widely used format in major sports, including the FIFA World Cup group stage until 2022. Both approaches are found to be highly relevant for historic FIFA World Cups, but the probabilistic model is shown to give an enhanced picture, especially by breaking down the broad category of competitive matches in the deterministic model.

Consequently, the probabilistic model of \citet{CsatoGyimesi2026b} is adopted to analyse the fundamental change in the design of the 2026 FIFA World Cup, which provides our second contribution. We decompose the overall effect of the reform, which is important because three independent rules have been modified at the same time:
(1) the number of participating teams has increased from 32 to 48;
(2) a higher proportion of teams (2/3 instead of 1/2) qualify from the groups since, besides the group winners and the runners-up, two-thirds of the third-placed teams advance to the knockout stage, too; and
(3) head-to-head records have replaced goal difference as the primary tie-breaking criteria in the groups.

The current study extends and refines the results of the pioneering work \citet{ChaterArrondelGayantLaslier2021}, on the competitiveness of the games played in the last round of the FIFA World Cup group stage, from several aspects. First, we use a more sophisticated, probabilistic categorisation of matches. In addition, the higher strength differences are explicitly accounted for in the simulation model, and the effect of the tie-breaking rules is also considered. Crucially, our attention is not restricted to the last group to play; averages are reported across all groups, addressing the challenge that the third-placed teams in the groups finishing earlier do not necessarily know the exact target of qualification. Finally, the decomposition procedure presented here helps in understanding the impact of each component in the 2026 reform of the FIFA World Cup.

The structure of the paper is as follows.
Section~\ref{Sec2} gives a concise overview of connected studies. The methodology is presented in Section~\ref{Sec3}. The deterministic and probabilistic classification models are compared in Section~\ref{Sec4}. Section~\ref{Sec5} uses the latter model to explore the effects of the recent comprehensive reform on the design of the FIFA World Cup. Section~\ref{Sec6} concludes.

\section{Related literature} \label{Sec2}

Applications of operational research to sports have grown rapidly in recent decades due to the huge amount of publicly available data, as well as the continuous development of computing techniques and statistical methods. This literature is surveyed by \citet{Wright2009, Wright2014, KendallLenten2017, Csato2021a, LentenKendall2022, DevriesereCsatoGoossens2025, DevriesereGoossensWillem2026}, among others. Here, we give an overview of two sets of studies that are relevant to our topic: works on the clustering of games with respect to the incentives of opposing teams, and research on the expansion of the FIFA World Cup.

Match classification is an increasingly popular topic in tournament design.
\citet{RibeiroUrrutia2005} present integer programming models that give necessary and sufficient conditions to ensure the qualification of a team in the first $m$ positions. Their technique can be used to detect in advance when a team has already qualified or been eliminated.
\citet{ChaterArrondelGayantLaslier2021} divide the matches played in the last round of the FIFA World Cup group stage into three distinct classes:
(a) competitive game if no team is indifferent and at least one outcome of the simultaneously played games threatens the qualification of one of them;
(b) collusive game if a particular outcome ensures the qualification of both teams;
(c) stakeless game if at least one team is indifferent because it is already qualified or eliminated.
The probabilities are computed via simulations for several reasonable formats with 32 teams (8 groups of four, top two qualify), 40 teams (8 groups of five, top two qualify), 48 teams (16 groups of three, top two qualify; 12 groups of four, top two plus the eight best third-placed teams qualify), and 60 teams (12 groups of five, top two plus the eight best third-placed teams qualify) under all possible schedules of the matches.

\citet{CsatoMolontayPinter2024} conduct a similar simulation for the UEFA Champions League groups, where four teams played a double round-robin tournament. However, the three categories of games are
(a) weakly stakeless if exactly one team is indifferent;
(b) strongly stakeless if both teams are indifferent;
(c) competitive otherwise.
Their probabilities are computed for both the penultimate and the last matchdays under the 12 valid schedules.
The same classification is used by \citet{Gyimesi2024} to compare the group and incomplete round-robin league formats of the UEFA Champions League, as well as the planned format of the European Super League.

\citet{DevriesereGoossensSpieksma2026} unify the classification models of \citet{ChaterArrondelGayantLaslier2021} and \citet{CsatoMolontayPinter2024} by considering competitive, asymmetric (weakly stakeless), (strongly) stakeless, and collusive games. Analogous to \citet{Gyimesi2024}, the group and incomplete round-robin formats of the UEFA Champions League are analysed, but not only random schedules are studied: the 12 valid schedules are examined for the group stage, and five reasonable schedules are generated for the incomplete round-robin league.

\citet{CsatoGyimesi2026a} refine the concept of stakeless games by taking into account their expected outcome. In particular, the organiser is assumed to face higher reputational costs if the indifferent team is more likely to win if it exerts full effort. A novel format with imbalanced groups is proposed for the 2026 FIFA World Cup to substantially decrease the probability that the 16 strongest teams become indifferent in the last round of the group stage.

Compared with these traditional models that use binary concepts, \citet{CsatoGyimesi2026b} propose a probabilistic classification model based on a risk-benefit analysis. First, a team is called indifferent if its payoff is fixed with a sufficiently high probability, rather than in any mathematically possible scenario. Second, a team is assumed to play more offensively/defensively if its expected gain from scoring goals is higher/lower than its expected loss from conceding goals. This implies six different types of matches, as will be detailed in Section~\ref{Sec312}.

It is known that multi-stage tournaments can always lead to strategic manipulation if more than one team qualifies from a group \citep{Vong2017}. However, allowing only one team to qualify makes teams indifferent with a high probability towards the end of the tournament, resulting in a number of asymmetric and stakeless games, which threatens the attractiveness and even the integrity of sports.
Thus, the investigation of trade-offs between desirable factors in tournament design calls for a probabilistic approach \citep{Vong2017}, as will be done here. 

Another line of literature considers and uses measures of match importance.
Based on the idea of \citet{Schilling1994}, \citet{ScarfShi2008} compute the impact of a match on the finishing positions, given the results of all other matches, some of them already played, some of them simulated. \citet{BuraimoForrestMcHaleTena2022} find that matches potentially significant for end-of-season outcomes attract a higher audience in the English Premier League. Compared to these studies, the main novelty of our match classification model is the focus on the two opposing teams, rather than on the match itself, which allows for quantifying team incentives.

This paper is also strongly connected to studies on various issues of the 2026 FIFA World Cup.
\citet{Guyon2020a} quantifies the risk of match fixing if the tournament is played in 16 groups of three, as was originally planned \citep{FIFA2017d}. Forcing the strongest team to play the first two matches minimises the risk of collusion, but this remains unacceptably high even in the most favourable case when all groups are strongly imbalanced. The problem cannot be avoided by forbidding draws. However, according to \citet{Stronka2024}, using random tie-breaking and dynamic scheduling can be effective: the expected number of games with a collusive outcome is reduced from 5.5 to 0.26.
These results have probably inspired FIFA to reconsider the design of the 2026 FIFA World Cup, and choose to organise it in 12 groups of four teams each, where the top two teams from each group and the eight best third-placed teams qualify for the Round of 32 \citep{FIFA2023b}. As we have seen, the competitiveness of matches in this format is analysed by \citet{ChaterArrondelGayantLaslier2021} and \citet{CsatoGyimesi2026a}.

The closest study is a parallel, not peer-reviewed work. \citet{Medina2026} builds a Monte Carlo simulation model for the FIFA World Cup, which reveals that uncompetitive final group games nearly doubled between 2022 and 2026, mainly due to head-to-head tiebreakers and the qualification of third-laced teams. It is shown that teams assigned to later finishing have a stronger temptation to play for
an advancing result. Crucially, the incentives are reinforced by empirical observations: teams for which a draw is beneficial indeed play a draw more frequently, in contrast to zero-sum final group games. Finally, an alternative 48-team format is suggested, where only the group winners and runners-up qualify for the knockout stage, and the best group winners directly qualify for the Round of 16.

\citet{DiMattiaKrumer2026} examine the effects of reducing the number of matches in groups of four teams. They exploit a structural change in professional beach volleyball, where the number of rounds has been reduced to two in 2017. The reform impacts neither the probability of a favourite team winning a single match nor the efficacy of the tournament.

The expansion to 48 teams has received attention from the perspective of slot allocation, too. The FIFA World Cup qualification is organised by the six confederations of FIFA, whose national teams compete for a given number of berths; only two slots are available in an intercontinental playoff tournament. Therefore, ensuring a balanced allocation of slots across the confederations is crucial for the fairness of the qualification \citep{StoneRod2016, Csato2023c}.
\citet{KrumerMoreno-Ternero2023} explore the problem by using the standard tools of the fair allocation literature. The claims are determined by the strengths of the confederations. 10 different allocations are considered based on various assumptions on the measure of strength and the status quo. UEFA (Union of European Football Associations) is found to deserve more slots than assigned by FIFA.
\citet{CsatoKissSzadoczki2025} develop an Elo-based method inspired by the FIFA World Ranking to quantify the performance of five confederations in previous FIFA World Cups. Their results show that more European and South American teams need to play in the competition than allowed by the official policy.

\section{Methodology} \label{Sec3}

Section~\ref{Sec31} presents the two existing approaches for match classification. The pre-2026 and post-2026 formats of the FIFA World Cup are described in Section~\ref{Sec32}, and our simulation framework is discussed in Section~\ref{Sec33}.

\subsection{Match classification models} \label{Sec31}

As we have outlined in Sections~\ref{Sec1} and \ref{Sec2}, two different frameworks have been suggested for clustering games played in the last rounds of sports tournaments.
They are presented in Sections~\ref{Sec31} and \ref{Sec32}, respectively, assuming that only one prize---qualification to the next stage---exists, which is realistic in the case of the FIFA World Cup \citep{ChaterArrondelGayantLaslier2021, Guyon2022a, CsatoGyimesi2026a}.

\subsubsection{Deterministic model} \label{Sec311}

The evolution of the deterministic approach for match classification has already been discussed in Section~\ref{Sec2}. Currently, \citet{DevriesereGoossensSpieksma2026} consider the most refined categorisation, by unifying the models introduced by \citet{ChaterArrondelGayantLaslier2021} and \citet{CsatoMolontayPinter2024}. We will use this model, which contains the following four distinct categories:
\begin{itemize}
\item 
Stakeless match: the prizes of both teams do not depend on the outcomes of matches played simultaneously or later;

\item 
Asymmetric match: the prize of exactly one team does not depend on the outcomes of matches played simultaneously or later;

\item 
Collusive match: the prizes of both teams depend on the outcomes of matches played simultaneously or later, but a particular result guarantees the prize for both teams;

\item 
Competitive match: none of the above, that is, the prizes of both teams depend on the outcomes of matches played simultaneously or later, and no particular result guarantees the prize for both teams.
\end{itemize}
Illustrative examples from historic FIFA World Cups will be given in Section~\ref{Sec43}.

\subsubsection{Probabilistic model} \label{Sec312}

According to our knowledge, the only probabilistic match classification model has been introduced by \citet{CsatoGyimesi2026b}.
Let $P_{i}^{(h)}$ be the probability that team $i$ obtains the prize if the result of its match is $h$.

Compared to the benchmark of 0-0, the potential gain from playing offensively is
\[
\mathcal{G}_i = P_{i}^{(m\text{-}0)} - P_{i, k}^{(0\text{-}0)},
\]
as the attacking team scores $m$ goals if its tactic is successful.

Compared to the benchmark of 0-0, the potential loss from playing offensively is
\[
\mathcal{L}_i = P_{i, k}^{(0\text{-}0)} - P_{i}^{(0\text{-}m)},
\]
as the attacking team concedes $m$ goals if its tactic is unsuccessful.

The classification is based on comparing the values $\mathcal{G}_i$ and $\mathcal{L}_i$ for the two opponents. Naturally, $\mathcal{G}_i$ and $\mathcal{L}_i$ are computed by simulating the results of all matches played simultaneously or later. The value of $m$ is also determined by the simulation model, which ensures that a weaker (stronger) team is less (more) likely to win by more goals.

An indifference threshold $\mathcal{I}$ is used to ignore events with a small probability of occurring: a team is said to be indifferent if both $\mathcal{G}_i$ and $\mathcal{L}_i$ remain below $\mathcal{I}$.
Teams play offensively by default, and shift to a defensive tactic only if the payoff from avoiding a loss ($\mathcal{L}_i$) is higher than the payoff from winning ($\mathcal{G}_i$).

\citet{CsatoGyimesi2026b} have adopted the probabilistic model to analyse the UEFA Champions League, where two prizes are available. Consequently, $\mathcal{G}_i = \mathcal{L}_i$ if a win certainly provides the first prize for team $i$, while a loss leads to no prize. The probability of such an event is non-marginal in the group stage used until the 2023/24 season, since the first prize is the top two positions, and the second prize is the third position in a group of four. Therefore, in order to avoid the uncertain classification when $\mathcal{G}_i = \mathcal{L}_i$, \citet{CsatoGyimesi2026b} have assumed that teams play defensively only if $\mathcal{L}_i$ exceeds $\mathcal{G}_i$ by the ``margin of error'' $\mathcal{I}$. In the case of the FIFA World Cup, this technical adjustment is unnecessary due to the existence of only one prize. Hence, we can choose the natural direct comparison of $\mathcal{G}_i$ and $\mathcal{L}_i$.

These assumptions imply six distinct categories of games:
\begin{itemize}
\item
Unimportant match: both teams are indifferent if $\max \left\{ \mathcal{G}_i; \mathcal{L}_i; \mathcal{G}_j; \mathcal{L}_j \right\} \leq \mathcal{I}$;

\item 
Defensive asymmetric match: team $i$ is indifferent and team $j$ should play defensively if $\max \left\{ \mathcal{G}_i; \mathcal{L}_i \right\} \leq \mathcal{I}$ and $\max \left\{ \mathcal{G}_j; \mathcal{L}_j \right\} > \mathcal{I}$ such that $\mathcal{G}_j < \mathcal{L}_j$;

\item 
Offensive asymmetric match: team $i$ is indifferent and team $j$ should play offensively if $\max \left\{ \mathcal{G}_i; \mathcal{L}_i \right\} \leq \mathcal{I}$ and $\max \left\{ \mathcal{G}_j; \mathcal{L}_j \right\} > \mathcal{I}$ such that $\mathcal{G}_j \geq \mathcal{L}_j$;

\item 
Antagonistic match: neither team is indifferent, team $i$ should play offensively and team $j$ should play defensively if $\max \left\{ \mathcal{G}_i; \mathcal{L}_i \right\} > \mathcal{I}$ and $\max \left\{ \mathcal{G}_j; \mathcal{L}_j \right\} > \mathcal{I}$ such that $\mathcal{G}_i \geq \mathcal{L}_i$ and $\mathcal{G}_j < \mathcal{L}_j$;

\item 
Defensive match: neither team is indifferent, both teams $i$ and $j$ should play defensively if $\max \left\{ \mathcal{G}_i; \mathcal{L}_i \right\} > \mathcal{I}$ and $\max \left\{ \mathcal{G}_j; \mathcal{L}_j \right\} > \mathcal{I}$ such that $\mathcal{G}_i < \mathcal{L}_i$ and $\mathcal{G}_j < \mathcal{L}_j$;

\item 
Offensive match: neither team is indifferent, both teams $i$ and $j$ should play offensively if $\max \left\{ \mathcal{G}_i; \mathcal{L}_i \right\} > \mathcal{I}$ and $\max \left\{ \mathcal{G}_j; \mathcal{L}_j \right\} > \mathcal{I}$ such that $\mathcal{G}_i \geq \mathcal{L}_i$ and $\mathcal{G}_j \geq \mathcal{L}_j$.
\end{itemize}
We set $\mathcal{I} = 0.02$, that is, teams become indifferent if the difference in the probability of qualification with a loss and a win remains below 2\%.
Illustrative examples from historic FIFA World Cups will be given in Section~\ref{Sec43}.

This framework even allows quantifying the strength of incentives for each match.
Let $\omega_i$ be 0 if team $i$ is indifferent, and $\omega_i = \lvert \mathcal{G}_i - \mathcal{L}_i \rvert$ if $\max \left\{ \mathcal{G}_i, \mathcal{L}_i \right\} > \mathcal{I}$.
The strength of incentives equals
(a) $\kappa = 0$ in an unimportant game;
(b) $\kappa = 100 \cdot \max \{ \omega_i ; \omega_j \}$ in a defensive or offensive asymmetric game; and
(c) $\kappa = 100 \cdot \min \left\{ \omega_i ; \omega_j \right\}$ in an antagonistic/defensive/offensive game.
Hence, $0 \leq \kappa \leq 100$, and a higher value indicates that more powerful incentives are created for the non-indifferent team(s). 

To define intensity, we deliberately use the maximum for asymmetric matches and the minimum if no team is indifferent. If exactly one team is indifferent, it does not care about the outcome, and everything depends on what the other team wants to do. On the other hand, if no team is indifferent, the conservative assumption is that the behaviour of the team with weaker incentives will decide what happens on the field.

\subsection{Tournament designs of the FIFA World Cup} \label{Sec32}

The FIFA World Cup was organised in essentially the same format between 1998 and 2022. The 32 participants were allocated into eight groups of four teams by maximising inter-confederation games played in the group stage \citep{Csato2025c}. While the groups were intended to be balanced in order to ensure fairness, this was usually not achieved due to the misaligned seeding policies \citep{Guyon2015a, CeaDuranGuajardoSureSiebertZamorano2020, Csato2023d, LaprePalazzolo2023}. In the groups, single round-robin tournaments were played, with three matches for each team. The first two teams advanced to the Round of 16, the last two were eliminated.
The final group standings were determined according to the following criteria:
\begin{enumerate}
\item
Points obtained in all group matches;
\item
Goal difference in all group matches;
\item
Number of goals scored in all group matches;
\item
Points obtained in the matches played between the teams in question;
\item
Goal difference in the matches played between the teams in question;
\item
Number of goals scored in the matches played between the teams in question;
\item
Fair play points in all group matches;
\item
Drawing of lots.
\end{enumerate}

The 2026 edition of the FIFA World Cup has seen several changes to the group stage. First, the number of teams increased from 32 to 48, divided into 12 groups of four.
Second, the knockout stage starts with the Round of 32, contested by the 12 group winners, the 12 runners-up, and the eight best third-placed teams. The third-placed teams are ranked by the relevant tie-breaking criteria of the 32-team format in the given order (1, 2, 3, 7, 8).
Third, in the 2026 edition, head-to-head results are preferred to goal difference; the order of tie-breaking criteria is 1, 4, 5, 6, 2, 3, 7, with the 8th replaced by position in the official FIFA World Rankings.




\subsection{The simulation model} \label{Sec33}

As usual in the literature of tournament design \citep{ScarfYusofBilbao2009}, we use simulations since historical data reflect only a single realisation of several uncertain parameters, and the probabilistic match classification model calls for computing the probability of qualification. In particular, the group stages of the five FIFA World Cups between 2010 and 2026 are simulated using the actual participants and group allocations. Hence, four instances of the 32-team format and one example of the 48-team format are considered.

Individual match outcomes---the number of goals scored by the two teams---are simulated by a Poisson model \citep{Maher1982}, which was estimated by \citet{FootballRankings2020} based on thousands of matches between national teams. This is a well-established simulation method in the academic literature, used by \citet{Csato2022a, Csato2023d, Csato2023a, Csato2023c, Csato2025d, CsatoGyimesi2026a, Stronka2024}.
The strengths of the teams are measured by their World Football Elo Ratings (\url{https://www.eloratings.net/}). In particular, ratings on 1 April 2026 are employed for the 48-team format, and 1 June of the given year for the 32-team format, downloaded from \url{https://www.international-football.net/}. The model distinguishes between matches played on neutral and non-neutral grounds. Naturally, only the host countries play at home. Following World Football Elo Ratings and in line with the regression of \citet{FootballRankings2020}, the Elo ratings of the host(s) are increased by 100 to reflect home advantage.

The simulations are conducted at two levels.
First, we generate results for all matches played simultaneously or later than the given match 1000 times. This is the set of matches whose outcomes remain unknown to the opposing teams at the start of their match, but could affect their qualification from the group. The 1000 simulated sets of results are used to quantify the probability of qualification under 1000 wins of $m$-$0$, 1000 draws of $0$-$0$, and 1000 losses of $0$-$m$. The value of $m$ comes directly from a probability distribution, analogous to \citet{CsatoGyimesi2026b}: the number of goals scored by the loser is set to 0, and the number of goals scored by the winner is generated by the Poisson model (if it would be 0, a new value is drawn). Thus, it is not fixed for a given match but varies across the simulations such that the expected value of $m$ is a monotonic function of Elo rating difference.
This simple procedure will be followed in Section~\ref{Sec43} to classify the 16-16 matches played in the last round of the group stage in the FIFA World Cups between 2014 and 2022.

However, in order to reliably compare the two classification models (Section~\ref{Sec44}) and examine the effects of the 2026 reform (Section~\ref{Sec5}), we need to generate larger and more representative samples for both the 32-team and the 48-team formats, without fixing the results of the first two rounds in the group stage to the actual outcomes. Therefore, the second level refers to simulating past results, too. Again, 1000 sets of results are generated for these earlier matches, yielding 1000 hypothetical scenarios for each FIFA World Cup between 2010 and 2026.

Tournament organisers should make decisions based on this representative set of outcomes, rather than the observed results of the first two rounds in historic competitions. For instance, it would be a serious mistake to assume that the champion of 2022, Argentina, is always defeated by Saudi Arabia in the group stage.

Finally, as it will turn out in Section~\ref{Sec41}, the deterministic model requires only simulating the first two rounds of the group stage in the pre-2026 format, when the top two teams qualify from each group. However, this simplification does not hold for the post-2026 format since only some third-placed teams qualify, making the groups interdependent: in any given group, the results in the other groups could affect qualification.
Hence, for example, \citet{ChaterArrondelGayantLaslier2021} focus on the last group to play, where the third-ranked team exactly knows its target to qualify---but this worst-case analysis is disadvantageous to the post-2026 format of the FIFA World Cup.
Fortunately, the unequal treatment of third-placed teams does not cause any difficulty for the probabilistic match classification model because the results of groups finishing earlier can be simulated accordingly.
To conclude, using the deterministic approach would be either complicated or biased for the 2026 FIFA World Cup, which provides an additional argument for considering the probabilistic approach to analyse the effects of the reform package.

\section{Which match classification model is better for the FIFA World Cup?} \label{Sec4}

Section~\ref{Sec41} classifies matches according to the deterministic model for single round-robin tournaments with four teams if the top two teams qualify.
The differences between the deterministic and probabilistic approaches are illustrated in Section~\ref{Sec42}.
The 48 games played in the last round of the FIFA World Cup group stages from 2014 to 2022 are classified by both models in Section~\ref{Sec43}.
Finally, Section~\ref{Sec44} compares the deterministic and probabilistic match classification models through the example of the four FIFA World Cups organised between 2010 and 2022.

\subsection{Identification of deterministic match types} \label{Sec41}

\begin{table}[t!]
  \centering
  \caption{Deterministic match classification in a four-team single round-robin \\ tournament if two teams qualify and the primary tie-breaking rule is goal difference}
  \label{Table1}
    \rowcolors{1}{}{gray!20}
    \begin{tabularx}{1\textwidth}{l CCC CCC} \toprule \showrowcolors
    Scenario  & 1     & 2     & 3     & 4     & 5     & 6 \\ \bottomrule
    A     & 2     & 3     & 4     & 4     & 4     & 4 \\
    B     & 2     & 3     & 2     & 3     & 3     & 4 \\
    C     & 2     & 3     & 2     & 2     & 3     & 1 \\
    D     & 2     & 3     & 1     & 1     & 1     & 1 \\ \hline
    Draws & 4     & 0     & 3     & 2     & 1     & 2 \\ \toprule \hiderowcolors
    \multirow{2}[0]{*}{Games} & A-B   & A-B   & A-B   & A-D   & A-B   & A-C \\
          & C-D   & C-D   & C-D   & B-C   & C-D   & B-D \\ \bottomrule
    \end{tabularx}

\vspace{0.25cm}   
\begin{threeparttable}
    \rowcolors{1}{}{gray!20}
    \begin{tabularx}{1\textwidth}{l CCCCC CCCC} \toprule \showrowcolors
    Scenario  & 7     & 8     & 9     & 10    & 11    & 12    & 13    & 14    & 15 \\ \bottomrule
    A     & 4     & 4     & 4     & 6     & 6     & 6     & 6     & 6     & 6 \\
    B     & 4     & 4     & 4     & 2     & 3     & 3     & 3     & 4     & 6 \\
    C     & 1     & 2     & 3     & 1     & 1     & 3     & 3     & 1     & 0 \\
    D     & 1     & 0     & 0     & 1     & 1     & 0     & 0     & 0     & 0 \\ \hline
    Draws & 2     & 2     & 1     & 2     & 1     & 0     & 0     & 1     & 0 \\ \toprule \hiderowcolors
    \multirow{2}[0]{*}{Games} & A-B   & A-B   & A-C   & A-B   & A-C   & A-D   & A-B   & A-B   & A-B \\
          & C-D   & C-D   & B-D   & C-D   & B-D   & B-C   & C-D   & C-D   & C-D \\ \midrule
    Collusive & A-B   & ---   & ---   & ---   & ---   & ---   & ---   & ---   & --- \\
    \multirow{2}[0]{*}{Asymmetric} & \multirow{2}[0]{*}{---} & \multirow{2}[0]{*}{C-D} & \multirow{2}[0]{*}{B-D} & \multirow{2}[0]{*}{A-B} & \multirow{2}[0]{*}{A-C} & \multirow{2}[0]{*}{---} & \multirow{2}[0]{*}{---} & A-B   & \multirow{2}[0]{*}{---} \\
          &       &       &       &       &       &       &       & C-D   &  \\
    \multirow{2}[0]{*}{Stakeless} & \multirow{2}[0]{*}{---} & \multirow{2}[0]{*}{---} & \multirow{2}[0]{*}{---} & \multirow{2}[0]{*}{---} & \multirow{2}[0]{*}{---} & \multirow{2}[0]{*}{A-D} & \multirow{2}[0]{*}{---} & \multirow{2}[0]{*}{---} & A-B \\
          &       &       &       &       &       &       &       &       & C-D \\ \bottomrule
    \end{tabularx}
\begin{tablenotes} \footnotesize
\item
The rows A--D indicate the number of points scored by the four teams after two rounds, respectively.
\item
The row Draws presents the number of draws played in the first two rounds.
\item
The row Games shows the matches to be played in the last round.
\end{tablenotes}
\end{threeparttable}
\end{table}

Table~\ref{Table1} presents all possible scenarios in a single round-robin tournament with four teams before the last round. The classification of matches in the deterministic model is also provided for the rules of the 2022 FIFA World Cup, when the top two teams qualify, and the primary tie-breaking criterion is goal difference.

Regarding the distribution of points, 13 cases may occur when two rounds are finished. If the points are 6-3-3-0 or 4-4-1-1, two different scenarios exist with respect to the matches played in the last round. It is important to distinguish these scenarios for the deterministic match classification model. If the points are 6-3-3-0 such that the top team A with 6 points plays against the bottom team D with 0 points, their game is stakeless: team A is already qualified, and team D is eliminated. However, in the other case under points 6-3-3-0, both last round games are competitive. Analogously, if the points are 4-4-1-1 such that the top teams A and B with 4 points play against each other in the last round, their game is collusive: both teams are guaranteed to qualify with a (goalless) draw. On the other hand, both last-round games are competitive if a team with 4 points plays against a team with 1 point.

Among the 15 possible scenarios, only one (scenario 7 in Table~\ref{Table1}) involves a collusive game.
The number of asymmetric games is two in one scenario (14), and one in four scenarios (8--11).
There exists one scenario (15) with two stakeless games, and another one (12) with one stakeless game.
Finally, there is no scenario with two different types of uncompetitive (collusive/asymmetric/stakeless) games, and both last round matches are competitive in six scenarios (1--6 and 13).

\subsection{Importance of distinguishing competitive matches} \label{Sec42}

\begin{table}[t!]
\centering
\caption{2022 FIFA World Cup, Group C: Standing before the last round}
\label{Table2}

\begin{threeparttable}
\rowcolors{3}{}{gray!20}
    \begin{tabularx}{\linewidth}{Cl CCC CCC >{\bfseries}C} \toprule \hiderowcolors
    Pos   & Team  & W     & D     & L     & GF    & GA    & GD    & Pts \\ \bottomrule \showrowcolors
    1     & Poland & 1     & 1     & 0     & 2     & 0     & $+$2     & 4 \\
    2     & Argentina & 1     & 0     & 1     & 3     & 2     & $+$1     & 3 \\ \hline
    3     & Saudi Arabia & 1     & 0     & 1     & 2     & 3     & $-$1    & 3 \\
    4     & Mexico & 0     & 1     & 1     & 0     & 2     & $-$2    & 1 \\ \bottomrule    
    \end{tabularx}
    
    \begin{tablenotes} \footnotesize
\item
Pos = Position; W = Won; D = Drawn; L = Lost; GF = Goals for; GA = Goals against; GD = Goal difference; Pts = Points. All teams played two matches.
\item
The top two teams qualify for the Round of 16, and the bottom two teams are eliminated.
\item
Matches in the last round: Argentina vs Poland, Saudi Arabia vs Mexico.
    \end{tablenotes}
\end{threeparttable}
\end{table}

Table~\ref{Table2} shows the standing of Group C in the 2022 FIFA World Cup before the last (third) round, where Argentina plays against Poland and Saudi Arabia plays against Mexico. Obviously, any team can still qualify or be eliminated; hence, none of these matches is asymmetric or stakeless. In addition, Argentina and Poland have no collusive strategy since a win by Saudi Arabia eliminates one of them. Therefore, both matches played in the last round are competitive.

Nonetheless, both Argentina and Poland qualify by playing a goalless draw if Mexico does not lose or does not win by at least three goals against Saudi Arabia. This has a relatively high probability since Mexico is a stronger team than Saudi Arabia: in the FIFA Men's World Rankings of 31 March 2022, used for seeding the 2022 FIFA World Cup, Mexico was ranked 9th, while Saudi Arabia was ranked 49th.
On the other hand, Argentina is certainly eliminated if it loses. Poland is also eliminated if it loses and either Saudi Arabia wins in the other match, or Saudi Arabia plays a draw and Argentina wins by a high margin, or both Mexico and Argentina win by a cumulated margin of five goals.
Consequently, neither Argentina nor Poland should accept a substantial risk in order to win their last game.

The incentives in the other match are quite different. Mexico needs to win in order to have any chance to qualify. Even though Saudi Arabia may qualify by playing a draw if Poland wins, this is unlikely, as Argentina is stronger than Poland. Thus, both Saudi Arabia and Mexico have to take a high risk in order to win their last game and have a decent chance of qualification.

Thus, both matches played in the last round are in the same (competitive) category according to the standard deterministic match classification model. However, both teams have strong incentives to defend the result of 0-0 in the first match, and both teams have powerful incentives to attack in the second match.

\subsection{Classification of historic FIFA World Cup games} \label{Sec43}

\begin{table}[t!]
  \centering
  \caption{Classification of matches played in the last \\ round of the 2022 FIFA World Cup group stage}
  \label{Table3}
\centerline{
\begin{threeparttable}
    \rowcolors{1}{gray!20}{}
    \begin{tabularx}{1.15\textwidth}{cll CCCC CCC} \toprule \hiderowcolors
    \multirow{2}[0]{*}{Group} & \multirow{2}[0]{*}{Team 1} & \multirow{2}[0]{*}{Team 2} & \multirow{2}[0]{*}{$\mathcal{L}_1$(\%)} & \multirow{2}[0]{*}{$\mathcal{G}_1$(\%)} & \multirow{2}[0]{*}{$\mathcal{L}_2$(\%)} & \multirow{2}[0]{*}{$\mathcal{G}_2$(\%)} & \multirow{2}[0]{*}{$\kappa$} & \multicolumn{2}{c}{Model} \\
          &       &       &       &       &       &       &       & \multicolumn{1}{c}{Det.} & \multicolumn{1}{c}{Prob.} \\ \bottomrule \showrowcolors
    A     & Ecuador & Senegal & 91.94 & 0     & 2.39  & 97.61 & 91.94 & C     & AN \\
    A     & Qatar & Netherlands & 0     & 0     & 25.57 & 0     & 25.57 & AS    & DAS \\ \hline
    B     & England & Wales & 1.66  & 0     & 0     & 27.70  & 27.70  & C     & OAS \\
    B     & Iran  & United States & 85.24 & 14.76 & 0     & 100   & 70.48 & C     & AN \\ \hline
    C     & Argentina & Poland & 71.67 & 28.33 & 68.80  & 0     & 43.34 & C     & D \\
    C     & Saudi Arabia & Mexico & 25.67 & 74.33 & 0     & 57.59 & 48.66 & C     & O \\ \hline
    D     & Australia & Denmark & 98.34 & 1.66  & 0     & 99.62 & 96.68 & C     & AN \\
    D     & France & Tunisia & 0     & 0     & 0     & 23.56 & 23.56 & AS    & OAS \\ \hline
    E     & Costa Rica & Germany & 68.90  & 31.10  & 0     & 83.63 & 37.80 & C     & AN \\
    E     & Spain & Japan & 15.51 & 0     & 19.53 & 80.47 & 15.51 & C     & AN \\ \hline
    F     & Belgium & Croatia & 2.72  & 97.28 & 67.56 & 0     & 67.56 & C     & AN \\
    F     & Canada & Morocco & 0     & 0     & 55.20  & 0     & 55.20 & AS    & DAS \\ \hline
    G     & Brazil & Cameroon & 0     & 0     & 0     & 42.91 & 42.91 & AS    & OAS \\
    G     & Serbia & Switzerland & 0     & 99.37 & 98.98 & 1.02  & 97.96 & C     & AN \\ \hline
    H     & Ghana & Uruguay & 93.50  & 6.50   & 0     & 90.62 & 87.00 & C     & AN \\
    H     & Portugal & South Korea & 0     & 0     & 0     & 50.50  & 50.50 & AS    & OAS \\ \bottomrule
    \end{tabularx}
\begin{tablenotes} \footnotesize
\item
Models: Det.\ stands for Deterministic, Prob.\ stands for Probabilistic.
\item
Deterministic match types: AS = asymmetric; C = competitive; COL = collusive; S = stakeless.
\item
Probabilistic match types: AN = antagonistic; D = defensive; DAS = defensive asymmetric; O = offensive; OAS = offensive asymmetric; U = unimportant.
\end{tablenotes}
\end{threeparttable}
}
\end{table}

Table~\ref{Table3} classifies the 16 matches played in the last round of the 2022 FIFA World Cup group stage based on both approaches presented in Section~\ref{Sec31}.

We have argued in Section~\ref{Sec42} that the two matches played in Group C significantly differ with respect to the incentives of the teams, which is not reflected by the deterministic model, but is captured by the probabilistic model. Indeed, according to our simulations, Argentina and Poland qualify with probabilities of more than 70\% and 100\%, respectively, even by playing a goalless draw. However, if they lose, Argentina is certainly eliminated, and Poland is also eliminated with a probability of more than 65\%. Therefore, this match is categorised as defensive in the probabilistic model. The strength of incentives $\kappa$ is remarkably high; the difference between the loss from losing and the gain from winning exceeds 40 percentage points for both teams.

On the other hand, Mexico cannot qualify by playing a draw, and Saudi Arabia has a chance of less than 30\% to qualify if it plays a goalless draw. Hence, they face strong incentives to attack, implying an offensive match. The strength of incentives $\kappa$ is also quite high; the difference between the gain from winning and the loss from losing almost reaches 50 percentage points for both teams.

In the deterministic model, there are five asymmetric matches, either because one team is already qualified (France, Brazil, Portugal) or eliminated (Qatar, Canada) after two rounds, and 11 competitive matches.
Naturally, in the probabilistic model, the asymmetric games are either defensive or offensive asymmetric, depending on whether the non-indifferent team prefers avoiding a loss to winning (e.g.\ the Netherlands certainly qualifies by playing a goalless draw), or winning to avoiding a loss (e.g.\ Tunisia is certainly eliminated by playing a goalless draw).
Among the competitive matches, there is also one offensive asymmetric match: in Group B, England can still be eliminated if it loses, but this has only a marginal probability of less than 2\%, below the indifference threshold $\mathcal{I}$. Hence, the probabilistic framework identifies England as an indifferent team.
The remaining 10 competitive matches are all antagonistic, except for the one defensive and the one offensive match in Group C, discussed above. Nevertheless, the strength of incentives varies from about 15 percentage points (Spain qualifies with a probability of more than 84\% even if it loses against Japan) to almost 98 percentage points.

The matches played in the last round of the 2014 and 2018 FIFA World Cup group stages, respectively, are classified in Tables~\ref{Table_A1} and \ref{Table_A2} in the Appendix.
Two stakeless/unimportant matches appear in Group B in 2014 and in two groups (Groups A and G) in 2018, corresponding to scenario 15 in Table~\ref{Table1}.
Group D in the 2014 FIFA World Cup has one stakeless/unimportant and one competitive/antagonistic match. Indeed, Costa Rica with 6 points is already qualified, while its opponent, England, is certainly eliminated. In the game Uruguay vs Italy, two teams with 3 points play against each other, but their goal differences are $-1$ and $0$, favouring Italy in the case of a draw. Thus, the strength of incentives $\kappa$ reaches its theoretical maximum because Italy qualifies at the expense of Uruguay if the result is a draw, while Uruguay qualifies at the expense of Italy if it wins.
Finally, there is one collusive match in the 2014 FIFA World Cup, as Group G corresponds to scenario 7 in Table~\ref{Table1}: two teams with 4 points, Germany and the United States, play against each other, while the two remaining teams have only 1 point each.

To conclude, all four deterministic match types occurred in the 2014 FIFA World Cup, and all six probabilistic match types occurred in the 2018 FIFA World Cup. Thus, both classifications presented in Section~\ref{Sec31} have a high practical relevance; they are not only theoretical curiosities.

\subsection{Comparison of the alternative match classification models} \label{Sec44}

\begin{figure}[t!]
\centering

\begin{subfigure}{\textwidth}
	\caption{Deterministic model}
	\label{Fig1a}
\begin{tikzpicture}
\begin{axis}[
name = axis1,
width = 0.965\textwidth, 
height = 0.6\textwidth,
xmajorgrids,
ymajorgrids,
scaled x ticks = false,
xlabel = {Probability (\%)},
xlabel style = {align=center, font=\small},
xticklabel style = {/pgf/number format/fixed,/pgf/number format/precision=5},
xmin = 0,
xbar stacked,
bar width = 8pt,
symbolic y coords = {{2-2-2-2 (1)},{3-3-3-3 (2)},{4-2-2-1 (3)},{4-3-2-1 (4)},{4-3-3-1 (5)},{4-4-1-1 (6)},{4-4-1-1 (7)},{4-4-2-0 (8)},{4-4-3-0 (9)},{6-2-1-1 (10)},{6-3-1-1 (11)},{6-3-3-0 (12)},{6-3-3-0 (13)},{6-4-1-0 (14)},{6-6-0-0 (15)}},
ytick = data,
y dir = reverse,
enlarge y limits = 0.05,
legend style = {font=\small,at={(0.05,-0.15)},anchor=north west,legend columns=4},
legend entries = {Stakeless$\qquad$, Asymmetric$\qquad$, Collusive$\qquad$, Competitive},
]
\addplot [red, thick, pattern = vertical lines, pattern color = red] coordinates{
(0,{2-2-2-2 (1)})
(0,{3-3-3-3 (2)})
(0,{4-2-2-1 (3)})
(0,{4-3-2-1 (4)})
(0,{4-3-3-1 (5)})
(0,{4-4-1-1 (6)})
(0,{4-4-1-1 (7)})
(0,{4-4-2-0 (8)})
(0,{4-4-3-0 (9)})
(0,{6-2-1-1 (10)})
(0,{6-3-1-1 (11)})
(5.828125,{6-3-3-0 (12)})
(0,{6-3-3-0 (13)})
(0,{6-4-1-0 (14)})
(7.75,{6-6-0-0 (15)})
};
\addplot [blue, thick, pattern = dots, pattern color = blue] coordinates{
(0,{2-2-2-2 (1)})
(0,{3-3-3-3 (2)})
(0,{4-2-2-1 (3)})
(0,{4-3-2-1 (4)})
(0,{4-3-3-1 (5)})
(0,{4-4-1-1 (6)})
(0,{4-4-1-1 (7)})
(1.478125,{4-4-2-0 (8)})
(4.9640625,{4-4-3-0 (9)})
(1.6859375,{6-2-1-1 (10)})
(5.8796875,{6-3-1-1 (11)})
(0,{6-3-3-0 (12)})
(0,{6-3-3-0 (13)})
(12.7875,{6-4-1-0 (14)})
(0,{6-6-0-0 (15)})
};
\addplot [brown, thick, pattern = horizontal lines, pattern color = brown] coordinates{
(0,{2-2-2-2 (1)})
(0,{3-3-3-3 (2)})
(0,{4-2-2-1 (3)})
(0,{4-3-2-1 (4)})
(0,{4-3-3-1 (5)})
(0,{4-4-1-1 (6)})
(1.128125,{4-4-1-1 (7)})
(0,{4-4-2-0 (8)})
(0,{4-4-3-0 (9)})
(0,{6-2-1-1 (10)})
(0,{6-3-1-1 (11)})
(0,{6-3-3-0 (12)})
(0,{6-3-3-0 (13)})
(0,{6-4-1-0 (14)})
(0,{6-6-0-0 (15)})
};
\addplot [ForestGreen, thick, pattern = bricks, pattern color = ForestGreen] coordinates{
(0.175,{2-2-2-2 (1)})
(2.065625,{3-3-3-3 (2)})
(2.928125,{4-2-2-1 (3)})
(4.325,{4-3-2-1 (4)})
(6.5125,{4-3-3-1 (5)})
(2.8875,{4-4-1-1 (6)})
(1.128125,{4-4-1-1 (7)})
(1.478125,{4-4-2-0 (8)})
(4.9640625,{4-4-3-0 (9)})
(1.6859375,{6-2-1-1 (10)})
(5.8796875,{6-3-1-1 (11)})
(5.828125,{6-3-3-0 (12)})
(18.640625,{6-3-3-0 (13)})
(0,{6-4-1-0 (14)})
(0,{6-6-0-0 (15)})
};
\end{axis}
\end{tikzpicture}
\end{subfigure}

\vspace{0.5cm}
\begin{subfigure}{\textwidth}
	\caption{Probabilistic model}
	\label{Fig1b}
\begin{tikzpicture}
\begin{axis}[
name = axis1,
width = 0.965\textwidth, 
height = 0.6\textwidth,
xmajorgrids,
ymajorgrids,
scaled x ticks = false,
xlabel = {Probability (\%)},
xlabel style = {align=center, font=\small},
xticklabel style = {/pgf/number format/fixed,/pgf/number format/precision=5},
xmin = 0,
xbar stacked,
bar width = 8pt,
symbolic y coords = {{2-2-2-2 (1)},{3-3-3-3 (2)},{4-2-2-1 (3)},{4-3-2-1 (4)},{4-3-3-1 (5)},{4-4-1-1 (6)},{4-4-1-1 (7)},{4-4-2-0 (8)},{4-4-3-0 (9)},{6-2-1-1 (10)},{6-3-1-1 (11)},{6-3-3-0 (12)},{6-3-3-0 (13)},{6-4-1-0 (14)},{6-6-0-0 (15)}},
ytick = data,
y dir = reverse,
enlarge y limits = 0.05,
legend style = {font=\small,at={(0,-0.15)},anchor=north west,legend columns=3},
legend entries = {Unimportant$\qquad$, Defensive asymmetric$\qquad$, Offensive asymmetric, Defensive$\qquad \quad \:\:$, Offensive$\qquad \qquad \qquad \quad \:\:$, Antagonistic$\qquad \quad \:\:\:$},
]
\addplot [red, thick, pattern = vertical lines, pattern color = red] coordinates{
(0,{2-2-2-2 (1)})
(0,{3-3-3-3 (2)})
(0,{4-2-2-1 (3)})
(0.0640625,{4-3-2-1 (4)})
(0,{4-3-3-1 (5)})
(0.15625,{4-4-1-1 (6)})
(0.025,{4-4-1-1 (7)})
(0.003125,{4-4-2-0 (8)})
(0.7875,{4-4-3-0 (9)})
(0,{6-2-1-1 (10)})
(0.728125,{6-3-1-1 (11)})
(5.828125,{6-3-3-0 (12)})
(0,{6-3-3-0 (13)})
(1.459375,{6-4-1-0 (14)})
(7.75,{6-6-0-0 (15)})
};
\addplot [blue, thick, pattern = dots, pattern color = blue] coordinates{
(0,{2-2-2-2 (1)})
(0,{3-3-3-3 (2)})
(0,{4-2-2-1 (3)})
(0.0109375,{4-3-2-1 (4)})
(0,{4-3-3-1 (5)})
(0.0625,{4-4-1-1 (6)})
(0.0953125,{4-4-1-1 (7)})
(0,{4-4-2-0 (8)})
(4.1765625,{4-4-3-0 (9)})
(0,{6-2-1-1 (10)})
(0,{6-3-1-1 (11)})
(0,{6-3-3-0 (12)})
(3.60625,{6-3-3-0 (13)})
(5.715625,{6-4-1-0 (14)})
(0,{6-6-0-0 (15)})
};
\addplot [ForestGreen, thick, pattern = bricks, pattern color = ForestGreen] coordinates{
(0,{2-2-2-2 (1)})
(0,{3-3-3-3 (2)})
(0,{4-2-2-1 (3)})
(0.096875,{4-3-2-1 (4)})
(0,{4-3-3-1 (5)})
(0.084375,{4-4-1-1 (6)})
(0.0625,{4-4-1-1 (7)})
(1.478125,{4-4-2-0 (8)})
(0,{4-4-3-0 (9)})
(1.6859375,{6-2-1-1 (10)})
(5.1515625,{6-3-1-1 (11)})
(0,{6-3-3-0 (12)})
(3.2125,{6-3-3-0 (13)})
(5.6125,{6-4-1-0 (14)})
(0,{6-6-0-0 (15)})
};
\addplot [purple, thick, pattern = north east lines, pattern color = purple] coordinates{
(0,{2-2-2-2 (1)})
(0.01875,{3-3-3-3 (2)})
(0,{4-2-2-1 (3)})
(0,{4-3-2-1 (4)})
(0.575,{4-3-3-1 (5)})
(0,{4-4-1-1 (6)})
(1.021875,{4-4-1-1 (7)})
(0.3078125,{4-4-2-0 (8)})
(0.103125,{4-4-3-0 (9)})
(0,{6-2-1-1 (10)})
(0,{6-3-1-1 (11)})
(0,{6-3-3-0 (12)})
(1.103125,{6-3-3-0 (13)})
(0,{6-4-1-0 (14)})
(0,{6-6-0-0 (15)})
};
\addplot [green, thick, pattern = grid, pattern color = green] coordinates{
(0.0390625,{2-2-2-2 (1)})
(0.0484375,{3-3-3-3 (2)})
(0.6234375,{4-2-2-1 (3)})
(0.05625,{4-3-2-1 (4)})
(0.6078125,{4-3-3-1 (5)})
(0,{4-4-1-1 (6)})
(1.0515625,{4-4-1-1 (7)})
(0,{4-4-2-0 (8)})
(0,{4-4-3-0 (9)})
(1.1078125,{6-2-1-1 (10)})
(0.2140625,{6-3-1-1 (11)})
(0.003125,{6-3-3-0 (12)})
(1.1046875,{6-3-3-0 (13)})
(0,{6-4-1-0 (14)})
(0,{6-6-0-0 (15)})
};
\addplot [brown, thick, pattern = horizontal lines, pattern color = brown] coordinates{
(0.1359375,{2-2-2-2 (1)})
(1.9984375,{3-3-3-3 (2)})
(2.3046875,{4-2-2-1 (3)})
(4.096875,{4-3-2-1 (4)})
(5.3296875,{4-3-3-1 (5)})
(2.584375,{4-4-1-1 (6)})
(0,{4-4-1-1 (7)})
(1.1671875,{4-4-2-0 (8)})
(4.8609375,{4-4-3-0 (9)})
(0.578125,{6-2-1-1 (10)})
(5.665625,{6-3-1-1 (11)})
(5.825,{6-3-3-0 (12)})
(9.6140625,{6-3-3-0 (13)})
(0,{6-4-1-0 (14)})
(0,{6-6-0-0 (15)})
};
\end{axis}
\end{tikzpicture}
\end{subfigure}

\captionsetup{justification=centering}
\caption{Match classification in the last round of \\ simulated 2010--2022 FIFA World Cup group stages}
\label{Fig1}

\end{figure}
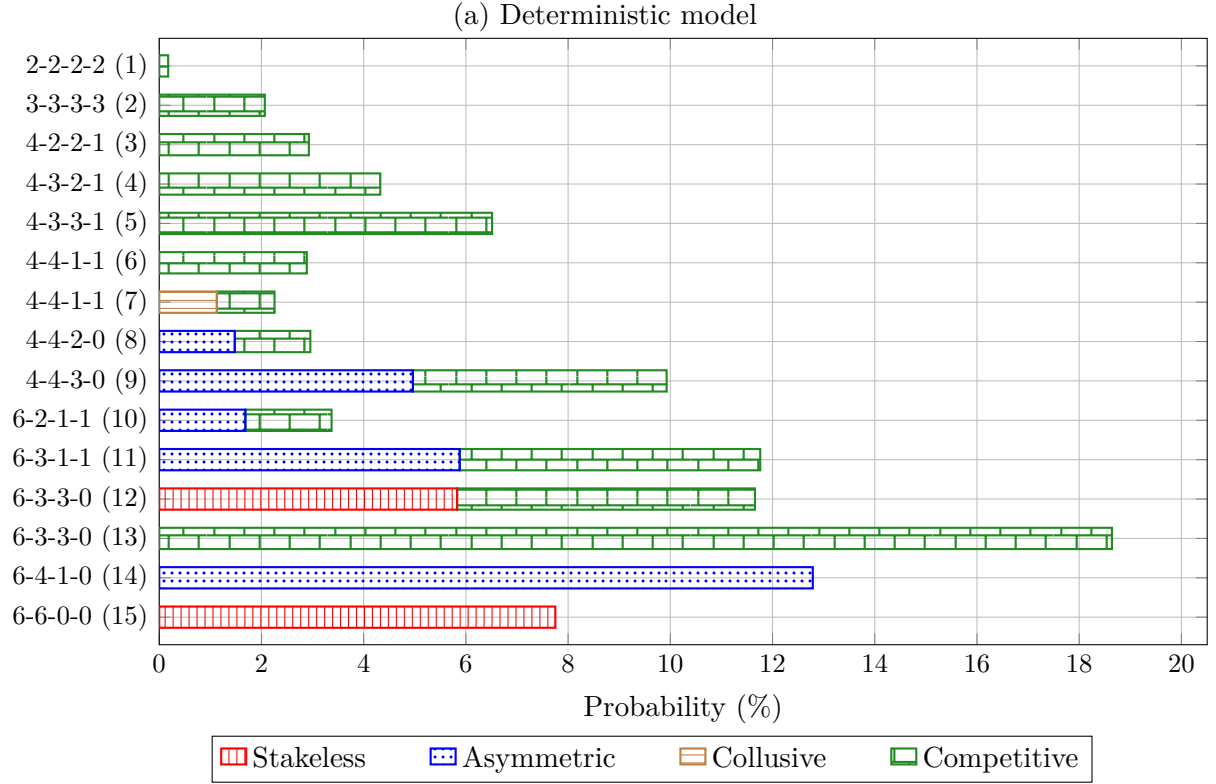
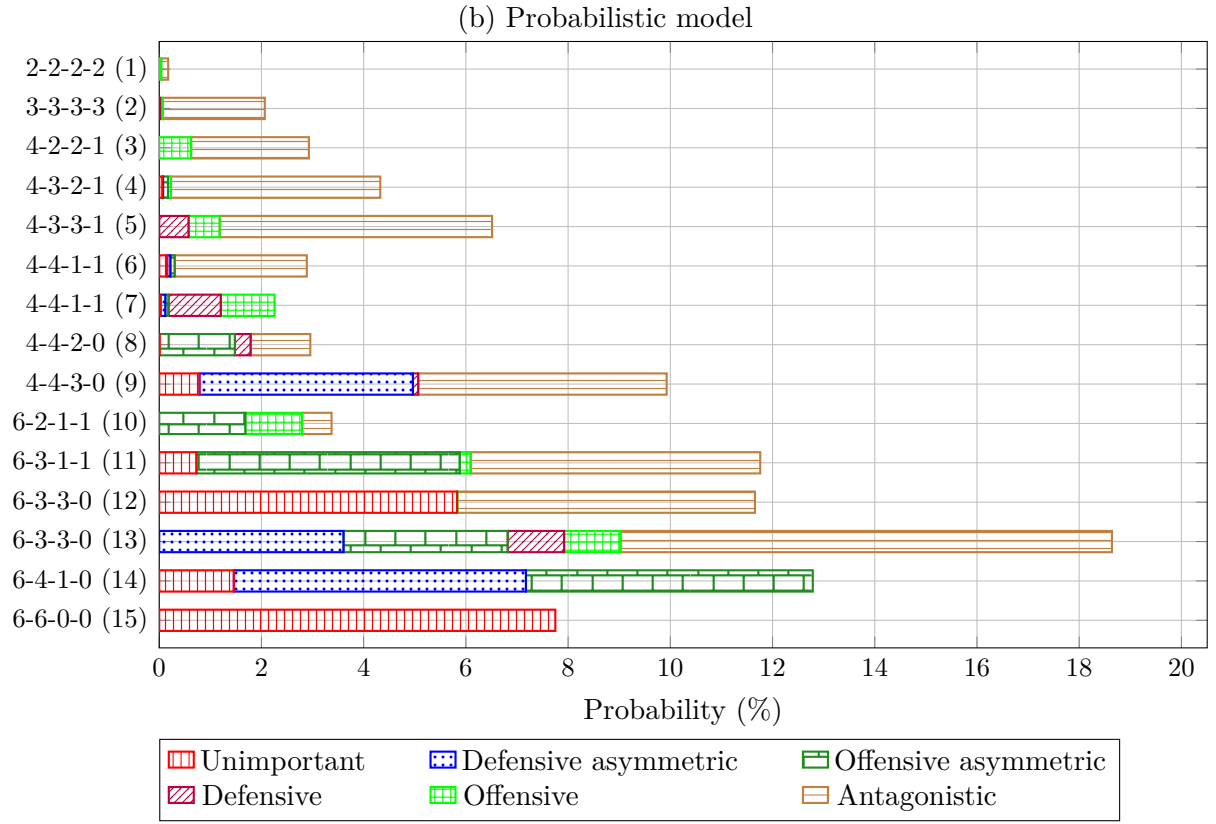


Figure~\ref{Fig1} plots the probability of each match type in the last round of the FIFA World Cup group stages from 2010 to 2022. All games played in the eight groups have been simulated 1000 times each, implying a sample of $4 \times 8 \times 2 \times 1000 = 64{,}000$.
Points 6-3-3-0 occur in about a third of simulated groups; the chances of scenarios 12 (when the two runners-up play against each other) and 13 (when the first team plays against a runner-up) are about 11.7\% and 18.6\%, respectively. The probability of three other cases (points 4-4-3-0, 6-3-1-1, and 6-4-1-0) approximates or exceeds 10\%, while it remains below 7\% in the further nine cases. Scenario 1, with four draws in the first two rounds, is almost impossible with a probability of merely 0.175\%.

Figure~\ref{Fig1a} for the deterministic model adds little to Table~\ref{Table1}. Competitive becomes far the most frequent category, but the proportion of asymmetric and stakeless matches is also substantial. On the other hand, collusive matches occur with a probability of less than 1.2\%. Given the standing after two rounds, at most two types of matches may emerge, as the last round contains two games.

Regarding the probabilistic model (Figure~\ref{Fig1b}), since a stakeless match is always unimportant, only unimportant games appear in scenario 15, but the picture is quite diverse in the majority of scenarios. All the six different types have a positive chance of occurring under points 4-4-1-1 and 6-3-3-0, which have two corresponding scenarios each. Antagonistic matches are dominant in five scenarios (1--5) when the group has a high competitive balance after two rounds. In line with the historical examples discussed in Section~\ref{Sec43}, offensive and, especially, defensive matches occur only occasionally, with probabilities of 4.9\% and 3.1\%, respectively.
Naturally, asymmetric matches mainly occur in imbalanced groups. Finally, besides scenario 15, a one-to-one correspondence exists between the deterministic and probabilistic models in scenario 12, where the competitive match played by the two teams with 3 points is always antagonistic.

\begin{table}[t!]
  \centering
  \caption{Frequency of deterministic and probabilistic match types in \\ the last round of simulated 2010--2022 FIFA World Cup group stages}
  \label{Table4}
  \rowcolors{1}{gray!20}{}
    \begin{tabularx}{\textwidth}{l cCCc c} \toprule \hiderowcolors
    \multirow{2}[0]{*}{Probabilistic model} & \multicolumn{4}{c}{Deterministic model} & \multirow{2}[0]{*}{Total} \\
          & Asymmetric & Collusive & Competitive & Stakeless & \\ \bottomrule \showrowcolors
    Antagonistic &       &       & 44.16\% &       & 44.16\% \\
    Defensive &       & 1.02\% & 2.11\% &       & 3.13\% \\
    Offensive &       &       & 4.86\% &       & 4.86\% \\
    Defensive asymmetric & 9.89\% & 0.10\% & 3.68\% &       & 13.67\% \\
    Offensive asymmetric & 13.93\% &       & 3.46\% &       & 17.38\% \\
    Unimportant & 2.98\% & 0.01\% & 0.24\% & 13.58\% & 16.80\% \\ \midrule \hiderowcolors
    Total & 26.80\% & 1.13\% & 58.50\% & 13.58\% & 100\% \\ \bottomrule
    \end{tabularx}
\end{table}

Table~\ref{Table4} compares the two match classification approaches for a characteristic FIFA World Cup with 32 teams, based on the average of the four tournaments played between 2010 and 2022.
Crucially, even though competitive matches have a probability of almost 60\%, they can be of any type in the probabilistic model, even though a majority of them are antagonistic. Stakeless matches are always unimportant, but some asymmetric matches become unimportant, too, if the non-indifferent team is almost certainly qualified or eliminated. The remaining asymmetric games are either defensive or offensive asymmetric. The joint probability of defensive and offensive asymmetric matches is higher than the probability of asymmetric matches in the deterministic model, because some competitive matches have an essentially guaranteed outcome for one of the opposing teams.
Even a collusive match may be unimportant if the two teams with 4 points have a sufficiently high goal difference after two rounds.
While most collusive games are defensive, the majority of defensive matches come from competitive matches.
Finally, antagonistic and offensive matches are always competitive.

To summarise, the two match classification models do differ to a substantial extent, as illustrated by Table~\ref{Table4}. The probabilistic model does not only refine the categories in the deterministic approach, but also gives an important insight into the incentives of the opposing teams.

\section{The effects of the 2026 reform package} \label{Sec5}

Now we evaluate the impact of the three main changes in the format of the 2026 FIFA World Cup group stage:
a) increasing the number of teams from 32 to 48 (Expansion effect);
b) allowing the qualification of two-thirds of the third-placed teams besides the group winners and the runners-up (Qualification effect); and
(c) preferring head-to-head results to goal difference in breaking ties (Tie-breaking effect).
The analysis is conducted by computing the probability of each type of game according to the recently developed, more sophisticated probabilistic match classification model.

All of these rule changes are expected to influence the distribution of matches.
First, the higher number of participants directly increases the differences between the strengths of the teams in all groups, especially because the allocation of slots among the FIFA confederations does not ensure the qualification of the best teams \citep{StoneRod2016, Csato2023c}, and this problem has been aggravated with 48 teams \citep{KrumerMoreno-Ternero2023}.
Second, when more teams qualify from a group, they are more (less) likely to qualify (be eliminated) before the last round---and the interplay of these opposite tendencies needs to be explored.
Third, favouring head-to-head records over goal difference may have non-negligible consequences with respect to the set of possible outcomes in the final group ranking \citep{Csato2023a, Csato2025d}.

\input{Figure2_reform_effect_match_category}

Figure~\ref{Fig2} presents the frequencies of the six match categories for the five FIFA World Cups between 2010 and 2026.
The proportion of unimportant games is doubled, from less than 20\% to almost 40\%, which might be worrying since both teams have few incentives to exert full effort in these games. In contrast, the ratio of defensive asymmetric matches appears to be similar in all editions of the FIFA World Cup. There is also a notable decrease in the frequency of offensive asymmetric games, where one team is indifferent, and the other has to focus on attacking.

Defensive matches have the lowest probability of occurrence in each case. Nonetheless, their current chance of around 4.7\% implies that at least one game out of the 24 played in the last round will be defensive with a probability of more than 70\%. A defensive match is expected to be boring as both opponents should defend rather than attack, and might threaten fairness by raising the suspicion of collusive behaviour, especially if a draw finally results in the qualification of both teams at the expense of a third one \citep{Guyon2020a}.

In contrast, the changes in the proportion of attractive offensive and antagonistic games are favourable. The former is at least tripled compared to the 32-team format, while the latter decreases from at least 40\% to less than 20\%. Note that offensive games are probably even more attractive than antagonistic games since both teams need to play offensively.

\begin{figure}[t!]
\centering

\begin{tikzpicture}
\begin{axis}[
name = axis1,
title = {32 teams, top two teams qualify, \\ goal difference},
title style = {font=\small,align=center},
width = 0.46\textwidth, 
height = 0.5\textwidth,
xmajorgrids,
ymajorgrids,
scaled x ticks = false,
xlabel = {Probability (\%)},
xlabel style = {align=center, font=\small},
xticklabel style = {/pgf/number format/fixed,/pgf/number format/precision=5},
xmin = 0,
xmax = 21,
xbar stacked,
bar width = 6pt,
symbolic y coords = {{2-2-2-2 (1)},{3-3-3-3 (2)},{4-2-2-1 (3)},{4-3-2-1 (4)},{4-3-3-1 (5)},{4-4-1-1 (6)},{4-4-1-1 (7)},{4-4-2-0 (8)},{4-4-3-0 (9)},{6-2-1-1 (10)},{6-3-1-1 (11)},{6-3-3-0 (12)},{6-3-3-0 (13)},{6-4-1-0 (14)},{6-6-0-0 (15)}},
ytick = data,
y dir = reverse,
enlarge y limits = 0.05,
]
\addplot [red, thick, pattern = vertical lines, pattern color = red] coordinates{
(0,{2-2-2-2 (1)})
(0,{3-3-3-3 (2)})
(0,{4-2-2-1 (3)})
(0.0640625,{4-3-2-1 (4)})
(0,{4-3-3-1 (5)})
(0.15625,{4-4-1-1 (6)})
(0.025,{4-4-1-1 (7)})
(0.003125,{4-4-2-0 (8)})
(0.7875,{4-4-3-0 (9)})
(0,{6-2-1-1 (10)})
(0.728125,{6-3-1-1 (11)})
(5.828125,{6-3-3-0 (12)})
(0,{6-3-3-0 (13)})
(1.459375,{6-4-1-0 (14)})
(7.75,{6-6-0-0 (15)})
};
\addplot [blue, thick, pattern = dots, pattern color = blue] coordinates{
(0,{2-2-2-2 (1)})
(0,{3-3-3-3 (2)})
(0,{4-2-2-1 (3)})
(0.0109375,{4-3-2-1 (4)})
(0,{4-3-3-1 (5)})
(0.0625,{4-4-1-1 (6)})
(0.0953125,{4-4-1-1 (7)})
(0,{4-4-2-0 (8)})
(4.1765625,{4-4-3-0 (9)})
(0,{6-2-1-1 (10)})
(0,{6-3-1-1 (11)})
(0,{6-3-3-0 (12)})
(3.60625,{6-3-3-0 (13)})
(5.715625,{6-4-1-0 (14)})
(0,{6-6-0-0 (15)})
};
\addplot [ForestGreen, thick, pattern = bricks, pattern color = ForestGreen] coordinates{
(0,{2-2-2-2 (1)})
(0,{3-3-3-3 (2)})
(0,{4-2-2-1 (3)})
(0.096875,{4-3-2-1 (4)})
(0,{4-3-3-1 (5)})
(0.084375,{4-4-1-1 (6)})
(0.0625,{4-4-1-1 (7)})
(1.478125,{4-4-2-0 (8)})
(0,{4-4-3-0 (9)})
(1.6859375,{6-2-1-1 (10)})
(5.1515625,{6-3-1-1 (11)})
(0,{6-3-3-0 (12)})
(3.2125,{6-3-3-0 (13)})
(5.6125,{6-4-1-0 (14)})
(0,{6-6-0-0 (15)})
};
\addplot [purple, thick, pattern = north east lines, pattern color = purple] coordinates{
(0,{2-2-2-2 (1)})
(0.01875,{3-3-3-3 (2)})
(0,{4-2-2-1 (3)})
(0,{4-3-2-1 (4)})
(0.575,{4-3-3-1 (5)})
(0,{4-4-1-1 (6)})
(1.021875,{4-4-1-1 (7)})
(0.3078125,{4-4-2-0 (8)})
(0.103125,{4-4-3-0 (9)})
(0,{6-2-1-1 (10)})
(0,{6-3-1-1 (11)})
(0,{6-3-3-0 (12)})
(1.103125,{6-3-3-0 (13)})
(0,{6-4-1-0 (14)})
(0,{6-6-0-0 (15)})
};
\addplot [green, thick, pattern = grid, pattern color = green] coordinates{
(0.0390625,{2-2-2-2 (1)})
(0.0484375,{3-3-3-3 (2)})
(0.6234375,{4-2-2-1 (3)})
(0.05625,{4-3-2-1 (4)})
(0.6078125,{4-3-3-1 (5)})
(0,{4-4-1-1 (6)})
(1.0515625,{4-4-1-1 (7)})
(0,{4-4-2-0 (8)})
(0,{4-4-3-0 (9)})
(1.1078125,{6-2-1-1 (10)})
(0.2140625,{6-3-1-1 (11)})
(0.003125,{6-3-3-0 (12)})
(1.1046875,{6-3-3-0 (13)})
(0,{6-4-1-0 (14)})
(0,{6-6-0-0 (15)})
};
\addplot [brown, thick, pattern = horizontal lines, pattern color = brown] coordinates{
(0.1359375,{2-2-2-2 (1)})
(1.9984375,{3-3-3-3 (2)})
(2.3046875,{4-2-2-1 (3)})
(4.096875,{4-3-2-1 (4)})
(5.3296875,{4-3-3-1 (5)})
(2.584375,{4-4-1-1 (6)})
(0,{4-4-1-1 (7)})
(1.1671875,{4-4-2-0 (8)})
(4.8609375,{4-4-3-0 (9)})
(0.578125,{6-2-1-1 (10)})
(5.665625,{6-3-1-1 (11)})
(5.825,{6-3-3-0 (12)})
(9.6140625,{6-3-3-0 (13)})
(0,{6-4-1-0 (14)})
(0,{6-6-0-0 (15)})
};
\end{axis}

\begin{axis}[
at = {(axis1.south east)},
xshift = 0.145\textwidth,
title = {48 teams, top two teams qualify, \\ goal difference},
title style = {font=\small,align=center},
width = 0.46\textwidth, 
height = 0.5\textwidth,
xmajorgrids,
ymajorgrids,
scaled x ticks = false,
xlabel = {Probability (\%)},
xlabel style = {align=center, font=\small},
xticklabel style = {/pgf/number format/fixed,/pgf/number format/precision=5},
xmin = 0,
xmax = 21,
xbar stacked,
bar width = 6pt,
symbolic y coords = {{2-2-2-2 (1)},{3-3-3-3 (2)},{4-2-2-1 (3)},{4-3-2-1 (4)},{4-3-3-1 (5)},{4-4-1-1 (6)},{4-4-1-1 (7)},{4-4-2-0 (8)},{4-4-3-0 (9)},{6-2-1-1 (10)},{6-3-1-1 (11)},{6-3-3-0 (12)},{6-3-3-0 (13)},{6-4-1-0 (14)},{6-6-0-0 (15)}},
ytick = data,
y dir = reverse,
enlarge y limits = 0.05,
]
\addplot [red, thick, pattern = vertical lines, pattern color = red] coordinates{
(0,{2-2-2-2 (1)})
(0,{3-3-3-3 (2)})
(0,{4-2-2-1 (3)})
(0.0791666666666667,{4-3-2-1 (4)})
(0,{4-3-3-1 (5)})
(0.179166666666667,{4-4-1-1 (6)})
(0.0583333333333333,{4-4-1-1 (7)})
(0,{4-4-2-0 (8)})
(0.5,{4-4-3-0 (9)})
(0,{6-2-1-1 (10)})
(0.4625,{6-3-1-1 (11)})
(3.85416666666667,{6-3-3-0 (12)})
(0,{6-3-3-0 (13)})
(4.15833333333333,{6-4-1-0 (14)})
(19.9666666666667,{6-6-0-0 (15)})
};
\addplot [blue, thick, pattern = dots, pattern color = blue] coordinates{
(0,{2-2-2-2 (1)})
(0,{3-3-3-3 (2)})
(0,{4-2-2-1 (3)})
(0,{4-3-2-1 (4)})
(0,{4-3-3-1 (5)})
(0.0583333333333333,{4-4-1-1 (6)})
(0.0875,{4-4-1-1 (7)})
(0,{4-4-2-0 (8)})
(3.7875,{4-4-3-0 (9)})
(0,{6-2-1-1 (10)})
(0,{6-3-1-1 (11)})
(0,{6-3-3-0 (12)})
(4.41666666666667,{6-3-3-0 (13)})
(6.77083333333333,{6-4-1-0 (14)})
(0,{6-6-0-0 (15)})
};
\addplot [ForestGreen, thick, pattern = bricks, pattern color = ForestGreen] coordinates{
(0,{2-2-2-2 (1)})
(0,{3-3-3-3 (2)})
(0,{4-2-2-1 (3)})
(0.0583333333333333,{4-3-2-1 (4)})
(0,{4-3-3-1 (5)})
(0.0625,{4-4-1-1 (6)})
(0.133333333333333,{4-4-1-1 (7)})
(1.40416666666667,{4-4-2-0 (8)})
(0,{4-4-3-0 (9)})
(1.30416666666667,{6-2-1-1 (10)})
(3.3875,{6-3-1-1 (11)})
(0,{6-3-3-0 (12)})
(4.44166666666667,{6-3-3-0 (13)})
(6.62083333333333,{6-4-1-0 (14)})
(0,{6-6-0-0 (15)})
};
\addplot [purple, thick, pattern = north east lines, pattern color = purple] coordinates{
(0,{2-2-2-2 (1)})
(0.025,{3-3-3-3 (2)})
(0,{4-2-2-1 (3)})
(0,{4-3-2-1 (4)})
(0.404166666666667,{4-3-3-1 (5)})
(0,{4-4-1-1 (6)})
(0.8375,{4-4-1-1 (7)})
(0.454166666666667,{4-4-2-0 (8)})
(0.0458333333333333,{4-4-3-0 (9)})
(0,{6-2-1-1 (10)})
(0,{6-3-1-1 (11)})
(0,{6-3-3-0 (12)})
(0.691666666666667,{6-3-3-0 (13)})
(0,{6-4-1-0 (14)})
(0,{6-6-0-0 (15)})
};
\addplot [green, thick, pattern = grid, pattern color = green] coordinates{
(0.0166666666666667,{2-2-2-2 (1)})
(0.0333333333333333,{3-3-3-3 (2)})
(0.433333333333333,{4-2-2-1 (3)})
(0.0333333333333333,{4-3-2-1 (4)})
(0.508333333333333,{4-3-3-1 (5)})
(0,{4-4-1-1 (6)})
(0.791666666666667,{4-4-1-1 (7)})
(0,{4-4-2-0 (8)})
(0,{4-4-3-0 (9)})
(0.908333333333333,{6-2-1-1 (10)})
(0.0708333333333333,{6-3-1-1 (11)})
(0.00833333333333333,{6-3-3-0 (12)})
(0.920833333333333,{6-3-3-0 (13)})
(0,{6-4-1-0 (14)})
(0,{6-6-0-0 (15)})
};
\addplot [brown, thick, pattern = horizontal lines, pattern color = brown] coordinates{
(0.0833333333333333,{2-2-2-2 (1)})
(1.21666666666667,{3-3-3-3 (2)})
(1.54166666666667,{4-2-2-1 (3)})
(2.87916666666667,{4-3-2-1 (4)})
(3.4375,{4-3-3-1 (5)})
(1.75,{4-4-1-1 (6)})
(0,{4-4-1-1 (7)})
(0.95,{4-4-2-0 (8)})
(4.24166666666667,{4-4-3-0 (9)})
(0.395833333333333,{6-2-1-1 (10)})
(3.77916666666667,{6-3-1-1 (11)})
(3.84583333333333,{6-3-3-0 (12)})
(7.90416666666667,{6-3-3-0 (13)})
(0,{6-4-1-0 (14)})
(0,{6-6-0-0 (15)})
};
\end{axis}
\end{tikzpicture}

\vspace{0.5cm}
\begin{tikzpicture}
\begin{axis}[
name = axis1,
title = {48 teams, the best third-placed \\ teams also qualify, goal difference},
title style = {font=\small,align=center},
width = 0.46\textwidth, 
height = 0.5\textwidth,
xmajorgrids,
ymajorgrids,
scaled x ticks = false,
xlabel = {Probability (\%)},
xlabel style = {align=center, font=\small},
xticklabel style = {/pgf/number format/fixed,/pgf/number format/precision=5},
xmin = 0,
xmax = 21,
xbar stacked,
bar width = 6pt,
symbolic y coords = {{2-2-2-2 (1)},{3-3-3-3 (2)},{4-2-2-1 (3)},{4-3-2-1 (4)},{4-3-3-1 (5)},{4-4-1-1 (6)},{4-4-1-1 (7)},{4-4-2-0 (8)},{4-4-3-0 (9)},{6-2-1-1 (10)},{6-3-1-1 (11)},{6-3-3-0 (12)},{6-3-3-0 (13)},{6-4-1-0 (14)},{6-6-0-0 (15)}},
ytick = data,
y dir = reverse,
enlarge y limits = 0.05,
legend style = {font=\small,at={(0.1,-0.2)},anchor=north west,legend columns=3},
legend entries = {Unimportant$\qquad$, Defensive asymmetric$\qquad$, Offensive asymmetric, Defensive$\qquad \quad \:\:$, Offensive$\qquad \qquad \qquad \quad \:\:$, Antagonistic$\qquad \quad \:\:\:$},
]
\addplot [red, thick, pattern = vertical lines, pattern color = red] coordinates{
(0,{2-2-2-2 (1)})
(0,{3-3-3-3 (2)})
(0,{4-2-2-1 (3)})
(0,{4-3-2-1 (4)})
(0.308333333333333,{4-3-3-1 (5)})
(0,{4-4-1-1 (6)})
(0.95,{4-4-1-1 (7)})
(1.4,{4-4-2-0 (8)})
(2.9875,{4-4-3-0 (9)})
(0,{6-2-1-1 (10)})
(0,{6-3-1-1 (11)})
(1.225,{6-3-3-0 (12)})
(2.65416666666667,{6-3-3-0 (13)})
(8.77083333333333,{6-4-1-0 (14)})
(10.475,{6-6-0-0 (15)})
};
\addplot [blue, thick, pattern = dots, pattern color = blue] coordinates{
(0,{2-2-2-2 (1)})
(0,{3-3-3-3 (2)})
(0.979166666666667,{4-2-2-1 (3)})
(0,{4-3-2-1 (4)})
(1.85833333333333,{4-3-3-1 (5)})
(0.0208333333333333,{4-4-1-1 (6)})
(0,{4-4-1-1 (7)})
(0.1625,{4-4-2-0 (8)})
(3.275,{4-4-3-0 (9)})
(1.29166666666667,{6-2-1-1 (10)})
(0.0666666666666667,{6-3-1-1 (11)})
(1.15,{6-3-3-0 (12)})
(8.45,{6-3-3-0 (13)})
(0.00416666666666667,{6-4-1-0 (14)})
(0,{6-6-0-0 (15)})
};
\addplot [ForestGreen, thick, pattern = bricks, pattern color = ForestGreen] coordinates{
(0,{2-2-2-2 (1)})
(0,{3-3-3-3 (2)})
(0.00833333333333333,{4-2-2-1 (3)})
(1.525,{4-3-2-1 (4)})
(0,{4-3-3-1 (5)})
(1.1625,{4-4-1-1 (6)})
(0,{4-4-1-1 (7)})
(0.0166666666666667,{4-4-2-0 (8)})
(2.29583333333333,{4-4-3-0 (9)})
(0.0125,{6-2-1-1 (10)})
(3.78333333333333,{6-3-1-1 (11)})
(2.925,{6-3-3-0 (12)})
(0.525,{6-3-3-0 (13)})
(0.891666666666667,{6-4-1-0 (14)})
(1.67083333333333,{6-6-0-0 (15)})
};
\addplot [purple, thick, pattern = north east lines, pattern color = purple] coordinates{
(0.1,{2-2-2-2 (1)})
(1.275,{3-3-3-3 (2)})
(0,{4-2-2-1 (3)})
(1.34166666666667,{4-3-2-1 (4)})
(0.00833333333333333,{4-3-3-1 (5)})
(0.0125,{4-4-1-1 (6)})
(0.00416666666666667,{4-4-1-1 (7)})
(0,{4-4-2-0 (8)})
(0.0125,{4-4-3-0 (9)})
(0.00416666666666667,{6-2-1-1 (10)})
(0.0625,{6-3-1-1 (11)})
(2.40416666666667,{6-3-3-0 (12)})
(0,{6-3-3-0 (13)})
(0,{6-4-1-0 (14)})
(0,{6-6-0-0 (15)})
};
\addplot [green, thick, pattern = grid, pattern color = green] coordinates{
(0,{2-2-2-2 (1)})
(0,{3-3-3-3 (2)})
(0,{4-2-2-1 (3)})
(0,{4-3-2-1 (4)})
(0,{4-3-3-1 (5)})
(0,{4-4-1-1 (6)})
(0.929166666666667,{4-4-1-1 (7)})
(0.00416666666666667,{4-4-2-0 (8)})
(0,{4-4-3-0 (9)})
(0.920833333333333,{6-2-1-1 (10)})
(0,{6-3-1-1 (11)})
(0,{6-3-3-0 (12)})
(0,{6-3-3-0 (13)})
(7.76666666666667,{6-4-1-0 (14)})
(7.82083333333333,{6-6-0-0 (15)})
};
\addplot [brown, thick, pattern = horizontal lines, pattern color = brown] coordinates{
(0,{2-2-2-2 (1)})
(0,{3-3-3-3 (2)})
(0.9875,{4-2-2-1 (3)})
(0.183333333333333,{4-3-2-1 (4)})
(2.175,{4-3-3-1 (5)})
(0.854166666666667,{4-4-1-1 (6)})
(0.025,{4-4-1-1 (7)})
(1.225,{4-4-2-0 (8)})
(0.00416666666666667,{4-4-3-0 (9)})
(0.379166666666667,{6-2-1-1 (10)})
(3.7875,{6-3-1-1 (11)})
(0.00416666666666667,{6-3-3-0 (12)})
(6.74583333333333,{6-3-3-0 (13)})
(0.116666666666667,{6-4-1-0 (14)})
(0,{6-6-0-0 (15)})
};
\end{axis}

\begin{axis}[
at = {(axis1.south east)},
xshift = 0.145\textwidth,
title = {48 teams, the best third-placed \\ teams also qualify, head-to-head},
title style = {font=\small,align=center},
width = 0.46\textwidth, 
height = 0.5\textwidth,
xmajorgrids,
ymajorgrids,
scaled x ticks = false,
xlabel = {Probability (\%)},
xlabel style = {align=center, font=\small},
xticklabel style = {/pgf/number format/fixed,/pgf/number format/precision=5},
xmin = 0,
xmax = 21,
xbar stacked,
bar width = 6pt,
symbolic y coords = {{2-2-2-2 (1)},{3-3-3-3 (2)},{4-2-2-1 (3)},{4-3-2-1 (4)},{4-3-3-1 (5)},{4-4-1-1 (6)},{4-4-1-1 (7)},{4-4-2-0 (8)},{4-4-3-0 (9)},{6-2-1-1 (10)},{6-3-1-1 (11)},{6-3-3-0 (12)},{6-3-3-0 (13)},{6-4-1-0 (14)},{6-6-0-0 (15)}},
ytick = data,
y dir = reverse,
enlarge y limits = 0.05,
]
\addplot [red, thick, pattern = vertical lines, pattern color = red] coordinates{
(0,{2-2-2-2 (1)})
(0,{3-3-3-3 (2)})
(0,{4-2-2-1 (3)})
(0,{4-3-2-1 (4)})
(0,{4-3-3-1 (5)})
(0,{4-4-1-1 (6)})
(0.95,{4-4-1-1 (7)})
(1.4,{4-4-2-0 (8)})
(5.75416666666667,{4-4-3-0 (9)})
(0,{6-2-1-1 (10)})
(0,{6-3-1-1 (11)})
(4.40416666666667,{6-3-3-0 (12)})
(2.62916666666667,{6-3-3-0 (13)})
(8.77083333333333,{6-4-1-0 (14)})
(10.4458333333333,{6-6-0-0 (15)})
};
\addplot [blue, thick, pattern = dots, pattern color = blue] coordinates{
(0,{2-2-2-2 (1)})
(0,{3-3-3-3 (2)})
(0.979166666666667,{4-2-2-1 (3)})
(0,{4-3-2-1 (4)})
(2.16666666666667,{4-3-3-1 (5)})
(0.0208333333333333,{4-4-1-1 (6)})
(0,{4-4-1-1 (7)})
(0.158333333333333,{4-4-2-0 (8)})
(2.80833333333333,{4-4-3-0 (9)})
(1.29166666666667,{6-2-1-1 (10)})
(0.0666666666666667,{6-3-1-1 (11)})
(1.39166666666667,{6-3-3-0 (12)})
(7.1875,{6-3-3-0 (13)})
(0.00416666666666667,{6-4-1-0 (14)})
(0,{6-6-0-0 (15)})
};
\addplot [ForestGreen, thick, pattern = bricks, pattern color = ForestGreen] coordinates{
(0,{2-2-2-2 (1)})
(0,{3-3-3-3 (2)})
(0.00833333333333333,{4-2-2-1 (3)})
(1.525,{4-3-2-1 (4)})
(0,{4-3-3-1 (5)})
(1.1625,{4-4-1-1 (6)})
(0,{4-4-1-1 (7)})
(0.00833333333333333,{4-4-2-0 (8)})
(0,{4-4-3-0 (9)})
(0.0125,{6-2-1-1 (10)})
(3.78333333333333,{6-3-1-1 (11)})
(0,{6-3-3-0 (12)})
(0,{6-3-3-0 (13)})
(0.825,{6-4-1-0 (14)})
(1.55,{6-6-0-0 (15)})
};
\addplot [purple, thick, pattern = north east lines, pattern color = purple] coordinates{
(0.1,{2-2-2-2 (1)})
(1.2625,{3-3-3-3 (2)})
(0,{4-2-2-1 (3)})
(1.35,{4-3-2-1 (4)})
(0.00833333333333333,{4-3-3-1 (5)})
(0.0125,{4-4-1-1 (6)})
(0.00416666666666667,{4-4-1-1 (7)})
(0,{4-4-2-0 (8)})
(0.0125,{4-4-3-0 (9)})
(0,{6-2-1-1 (10)})
(0.0625,{6-3-1-1 (11)})
(1.90833333333333,{6-3-3-0 (12)})
(0,{6-3-3-0 (13)})
(0,{6-4-1-0 (14)})
(0,{6-6-0-0 (15)})
};
\addplot [green, thick, pattern = grid, pattern color = green] coordinates{
(0,{2-2-2-2 (1)})
(0,{3-3-3-3 (2)})
(0,{4-2-2-1 (3)})
(0,{4-3-2-1 (4)})
(0,{4-3-3-1 (5)})
(0,{4-4-1-1 (6)})
(0.929166666666667,{4-4-1-1 (7)})
(0.00833333333333333,{4-4-2-0 (8)})
(0,{4-4-3-0 (9)})
(1.15416666666667,{6-2-1-1 (10)})
(0,{6-3-1-1 (11)})
(0,{6-3-3-0 (12)})
(0,{6-3-3-0 (13)})
(7.83333333333333,{6-4-1-0 (14)})
(7.97083333333333,{6-6-0-0 (15)})
};
\addplot [brown, thick, pattern = horizontal lines, pattern color = brown] coordinates{
(0,{2-2-2-2 (1)})
(0.0125,{3-3-3-3 (2)})
(0.9875,{4-2-2-1 (3)})
(0.175,{4-3-2-1 (4)})
(2.175,{4-3-3-1 (5)})
(0.854166666666667,{4-4-1-1 (6)})
(0.025,{4-4-1-1 (7)})
(1.23333333333333,{4-4-2-0 (8)})
(0,{4-4-3-0 (9)})
(0.15,{6-2-1-1 (10)})
(3.7875,{6-3-1-1 (11)})
(0.00416666666666667,{6-3-3-0 (12)})
(8.55833333333333,{6-3-3-0 (13)})
(0.116666666666667,{6-4-1-0 (14)})
(0,{6-6-0-0 (15)})
};
\end{axis}
\end{tikzpicture}

\captionsetup{justification=centering}
\caption{Probabilistic match classification in the \\ last round of different formats for the FIFA World Cup}
\label{Fig3}

\end{figure}
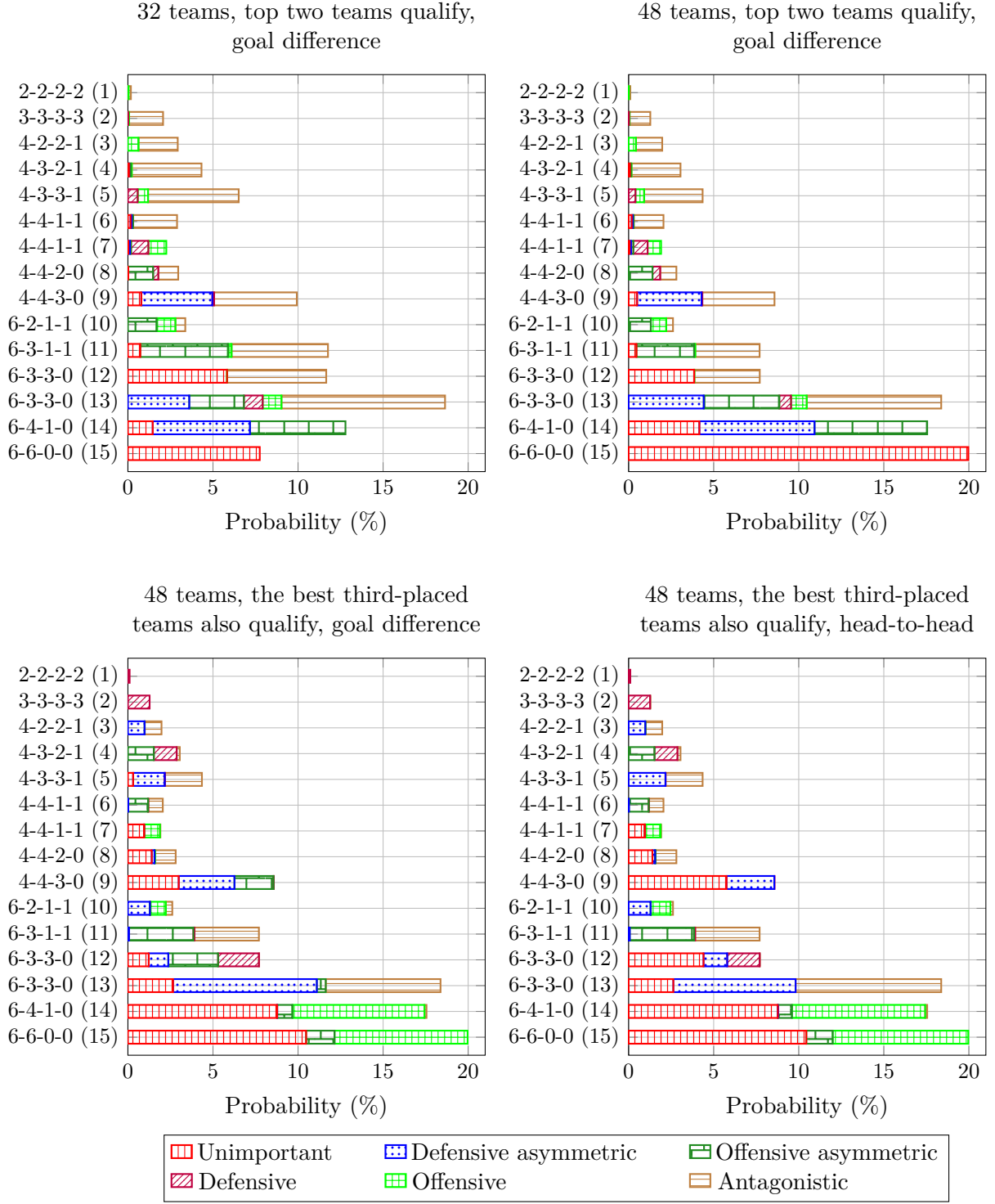


Figure~\ref{Fig3} details these changes by clustering the groups according to their standing before the last round (Table~\ref{Table1}), and reporting the results for four different formats:
\begin{itemize}
\item 
Standard 32-team format, used in the FIFA World Cup group stage until 2022 (top two teams qualify from each group, the primary tie-breaking rule is goal difference);
\item 
Expanded 48-team format (top two teams qualify from each group, the primary tie-breaking rule is goal difference);
\item
Expanded 48-team format with new qualification rules (top two teams qualify from each group together with the eight best third-placed teams, the primary tie-breaking rule is goal difference);
\item
Expanded 48-team format with new qualification and tie-breaking rules, used in the 2026 FIFA World Cup group stage (top two teams qualify from each group together with the eight best third-placed teams, the primary tie-breaking rule is head-to-head result).
\end{itemize}

The probability of the 15 cases is independent of the qualification and tie-breaking rules, but it is hugely influenced by the composition of teams. Increasing the number of teams to 48 strongly worsens competitive balance; for example, the chance that two teams win their first two matches (scenario 15) increases from 7.8\% to 20\%, which is mainly responsible for the higher frequency of unimportant matches.
However, the post-2026 qualification rules make most matches between the two teams with two losses each offensive or offensive asymmetric, since the winner can qualify among the best third-placed teams.
The tie-breaking rules have a non-negligible effect, too: they turn the majority of offensive asymmetric matches into unimportant ones in scenarios 9 and 12.

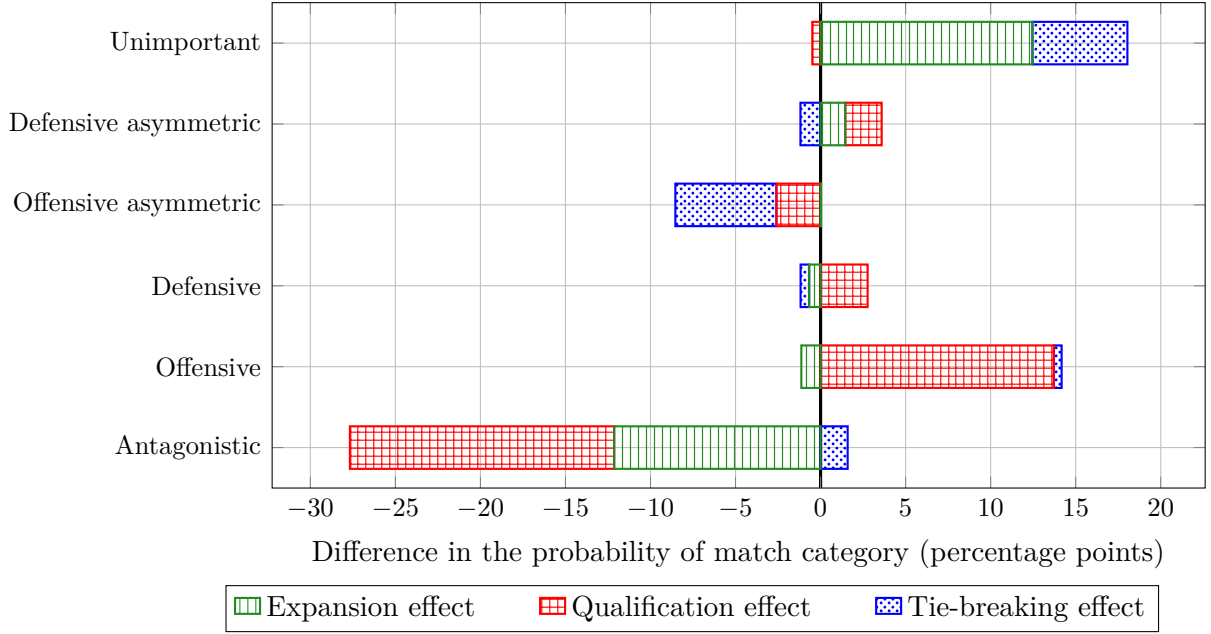
\begin{figure}[t!]
\centering

\begin{tikzpicture}
\begin{axis}[
width = 0.87\textwidth, 
height = 0.5\textwidth, 
xmajorgrids,
ymajorgrids,
xbar stacked,
bar width = 16pt,
scaled x ticks = false,
xlabel = {Difference in the probability of match category (percentage points)},
xlabel style = {align=center, font=\small},
xticklabel style = {/pgf/number format/fixed,/pgf/number format/precision=5},
extra x ticks = 0,
extra x tick labels = ,
extra x tick style = {grid = major, major grid style = {black,very thick}},
symbolic y coords = {Unimportant,Defensive asymmetric,Offensive asymmetric,Defensive,Offensive,Antagonistic},
ytick = data,
y dir = reverse,
enlarge y limits = 0.1,
legend style = {font=\small,at={(-0.05,-0.2)},anchor=north west,legend columns=3},
legend entries = {Expansion effect$\qquad$, Qualification effect$\qquad$, Tie-breaking effect},
]
\addplot [ForestGreen, thick, pattern = vertical lines, pattern color = ForestGreen] coordinates{
(12.4567708333333,Unimportant)
(1.45364583333333,Defensive asymmetric)
(0.0281250000000011,Offensive asymmetric)
(-0.671354166666666,Defensive)
(-1.13125,Offensive)
(-12.1359375,Antagonistic)
};
\addplot [red, thick, pattern = grid, pattern color = red] coordinates{
(-0.487500000000002,Unimportant)
(2.1375,Defensive asymmetric)
(-2.59583333333333,Offensive asymmetric)
(2.76666666666667,Defensive)
(13.7166666666667,Offensive)
(-15.5375,Antagonistic)
};
\addplot [blue, thick, pattern = crosshatch dots, pattern color = blue] coordinates{
(5.58333333333334,Unimportant)
(-1.18333333333333,Defensive asymmetric)
(-5.94166666666667,Offensive asymmetric)
(-0.504166666666667,Defensive)
(0.454166666666672,Offensive)
(1.59166666666667,Antagonistic)
};
\end{axis}
\end{tikzpicture}

\caption{Decomposition of changes in the distribution of probabilistic \\ match categories in the last round of the 2026 FIFA World Cup group stage}
\label{Fig4}

\end{figure}


Finally, Figure~\ref{Fig4} aggregates the detailed results reported in Figure~\ref{Fig3} and decomposes the changes in the distribution of match types in order to explore the implications of the three main effects separately. First, the expansion to 48 teams essentially shifts one-eighth of all games played in the last round from antagonistic to unimportant, which is clearly detrimental to attractiveness.
The easing of qualification reduces the proportion of antagonistic matches by even more, exceeding 15 percentage points. But the balance seems to be beneficial: the increase in the frequency of offensive matches greatly exceeds that of defensive matches. About 2-3\% of all last-round games are also moved from offensive asymmetric to defensive asymmetric, which is not welcome from the perspective of attractiveness.
Last but not least, tie-breaking rules affect mainly unimportant and offensive asymmetric matches, with an increase in the former category of about 5.6 percentage points. This is in line with previous results; preferring head-to-head results has been demonstrated to lead to more stakeless \citep{Csato2023a} and collusive \citep{Csato2025d} games in the deterministic model applied to round-robin contests among four teams.

\section{Concluding remarks} \label{Sec6}

The current paper has attempted to achieve two aims.
Our first, theoretical goal has been a better understanding of a recently developed probabilistic match classification model \citep{CsatoGyimesi2026b} by comparing it with the traditional deterministic approach \citep{RibeiroUrrutia2005, ChaterArrondelGayantLaslier2021, CsatoMolontayPinter2024, DevriesereGoossensSpieksma2026}. For this purpose, the format of a single round-robin group with four teams, where the first two teams qualify, has been chosen. In this simple but extensively used case, the number of possible scenarios before the last round remains manageable, and allows the identification of the deterministic match types without using integer programming.
By classifying the matches played in three historic FIFA World Cups, we have demonstrated the usefulness of the sophisticated probabilistic classification model, as all the six possible types of games have occurred in the 2018 competition. Furthermore, the probabilistic model has been shown to be fundamentally different from the deterministic model considered in the previous literature; for example, competitive matches can be of any type in our probabilistic framework.

Inspired by this promising clustering method, our second, practical goal has been exploring how the recent reform of the FIFA World Cup in 2026 has changed the distribution of matches. Since this has been a comprehensive package consisting of three main elements, we have decomposed the consequences of the increase in the number of teams, as well as the implications of the modified qualification and tie-breaking rules. Even though the participation of 16 weaker teams undeniably reduces the attractiveness of the last round played in the group stage by worsening competitive balance, allowing the qualification of the best third-placed teams is favourable for attractiveness due to exchanging antagonistic matches with offensive matches, where a draw is no longer advantageous for any team. In contrast, using head-to-head results as the primary criteria for breaking ties does not support competition due to restricting the set of possible group rankings after the first two rounds.

Our findings can be interesting for any tournament organiser planning to change the number of participants, the qualification system, or the tie-breaking principles, especially if the competition contains single round-robin groups of four teams. This format often appears in major sports, such as in the FIBA Basketball World Cup, the FIFA World Cup and the UEFA European Championship (in association football), the IHF World Handball Championship, or the FIVB Volleyball World Championship. We encourage governing bodies to carry out similar investigations in the future, even before similar reforms are introduced, and discuss with all stakeholders whether they can accept the expected consequences.

To summarise, this paper reinforces the validity of a recently proposed probabilistic approach to match classification and will probably inspire further studies that analyse the effects of rule changes in the design of sports competitions. Hopefully, it can also contribute to stronger collaboration between the academic community and decision-makers in sports.

\section*{Acknowledgements}
\addcontentsline{toc}{section}{Acknowledgements}

We are grateful to two anonymous colleagues for useful remarks. \\
The research was supported by the National Research, Development and Innovation Office under Grants Advanced 152220 and FK 145838, and the J\'anos Bolyai Research Scholarship of the Hungarian Academy of Sciences.

\bibliographystyle{apalike} 
\bibliography{All_references}

\clearpage
\setcounter{figure}{0}
\renewcommand{\thefigure}{A.\arabic{figure}}

\setcounter{table}{0}
\renewcommand{\thetable}{A.\arabic{table}}

\section*{Appendix}
\addcontentsline{toc}{section}{Appendix}

\begin{table}[ht!]
  \centering
  \caption{Classification of matches played in the last \\ round of the 2014 FIFA World Cup group stage}
  \label{Table_A1}
\centerline{
\begin{threeparttable}
    \rowcolors{1}{gray!20}{}
    \begin{tabularx}{1.15\textwidth}{cll CCCC CCC} \toprule \hiderowcolors
    \multirow{2}[0]{*}{Group} & \multirow{2}[0]{*}{Team 1} & \multirow{2}[0]{*}{Team 2} & \multirow{2}[0]{*}{$\mathcal{L}_1$(\%)} & \multirow{2}[0]{*}{$\mathcal{G}_1$(\%)} & \multirow{2}[0]{*}{$\mathcal{L}_2$(\%)} & \multirow{2}[0]{*}{$\mathcal{G}_2$(\%)} & \multirow{2}[0]{*}{$\kappa$} & \multicolumn{2}{c}{Model} \\
          &       &       &       &       &       &       &       & \multicolumn{1}{c}{Det.} & \multicolumn{1}{c}{Prob.} \\ \bottomrule \showrowcolors
    A     & Brazil & Cameroon & 26.10  & 0     & 0     & 0     & 26.10 & AS    & DAS \\
    A     & Croatia & Mexico & 0.79  & 99.21 & 100   & 0     & 98.42 & C     & AN \\ \hline
    B     & Netherlands & Chile & 0     & 0     & 0     & 0     & 0     & S     & U \\
    B     & Spain & Australia & 0     & 0     & 0     & 0     & 0     & S     & U \\ \hline
    C     & Colombia & Japan & 0     & 0     & 0     & 48.61 & 48.61 & AS    & OAS \\
    C     & Greece & Ivory Coast & 0     & 82.03 & 90.49 & 9.51  & 80.98 & C     & AN \\ \hline
    D     & Costa Rica & England & 0     & 0     & 0     & 0     & 0     & S     & U \\
    D     & Uruguay & Italy & 0     & 100   & 100   & 0     & 100   & C     & AN \\ \hline
    E     & Ecuador & France & 27.34 & 49.66 & 0.34  & 0     & 0.34  & C     & OAS \\
    E     & Switzerland & Honduras & 38.32 & 34.51 & 0     & 6.38  & 3.81  & C     & AN \\ \hline
    F     & Argentina & Nigeria & 0     & 0     & 28.05 & 0     & 28.05 & AS    & DAS \\
    F     & Bosnia and H. & Iran  & 0     & 0     & 0     & 65.24 & 65.24 & AS    & OAS \\ \hline
    G     & Germany & United States & 1.72  & 0     & 41.48 & 0     & 41.48  & COL   & DAS \\
    G     & Portugal & Ghana & 0     & 26.61 & 0     & 51.37 & 26.61 & C     & O \\ \hline
    H     & Algeria & Russia & 98.16 & 1.84  & 0     & 97.45 & 96.32 & C     & AN \\
    H     & Belgium & South Korea & 0     & 0     & 0     & 13.56 & 13.56 & AS    & OAS \\ \bottomrule
    \end{tabularx}
\begin{tablenotes} \footnotesize
\item
Teams: Bosnia and H.\ stands for Bosnia and Herzegovina.
\item
Models: Det.\ stands for Deterministic, Prob.\ stands for Probabilistic.
\item
Deterministic match types: AS = asymmetric; C = competitive; COL = collusive; S = stakeless.
\item
Probabilistic match types: AN = antagonistic; D = defensive; DAS = defensive asymmetric; O = offensive; OAS = offensive asymmetric; U = unimportant.
\end{tablenotes}
\end{threeparttable}
}
\end{table}

\begin{table}[t!]
  \centering
  \caption{Classification of matches played in the last \\ round of the 2018 FIFA World Cup group stage}
  \label{Table_A2}
\centerline{
\begin{threeparttable}
    \rowcolors{1}{gray!20}{}
    \begin{tabularx}{1.15\textwidth}{cll CCCC CCC} \toprule \hiderowcolors
    \multirow{2}[0]{*}{Group} & \multirow{2}[0]{*}{Team 1} & \multirow{2}[0]{*}{Team 2} & \multirow{2}[0]{*}{$\mathcal{L}_1$(\%)} & \multirow{2}[0]{*}{$\mathcal{G}_1$(\%)} & \multirow{2}[0]{*}{$\mathcal{L}_2$(\%)} & \multirow{2}[0]{*}{$\mathcal{G}_2$(\%)} & \multirow{2}[0]{*}{$\kappa$} & \multicolumn{2}{c}{Model} \\
          &       &       &       &       &       &       &       & \multicolumn{1}{c}{Det.} & \multicolumn{1}{c}{Prob.} \\ \bottomrule \showrowcolors
    A     & Russia & Uruguay & 0     & 0     & 0     & 0     & 0     & S     & U \\
    A     & Saudi Arabia & Egypt & 0     & 0     & 0     & 0     & 0     & S     & U \\ \hline
    B     & Portugal & Iran  & 98.01 & 0     & 1.86  & 98.14 & 96.28 & C     & AN \\
    B     & Spain & Morocco & 18.51 & 0     & 0     & 0     & 18.51 & AS    & DAS \\ \hline
    C     & Australia & Peru  & 0     & 51.21 & 0     & 0     & 51.21 & AS    & OAS \\
    C     & France & Denmark & 0     & 0     & 17.91 & 0     & 17.91 & AS    & DAS \\ \hline
    D     & Argentina & Nigeria & 0     & 86.48 & 88.13 & 11.87 & 76.26 & C     & AN \\
    D     & Iceland & Croatia & 0     & 39.63 & 0     & 0     & 39.63 & AS    & OAS \\ \hline
    E     & Brazil & Serbia & 81.38 & 0     & 8.47  & 91.53 & 81.38 & C     & AN \\
    E     & Switzerland & Costa Rica & 12.96 & 0     & 0     & 0     & 12.96 & AS    & DAS \\ \hline
    F     & Germany & South Korea & 48.17 & 46.58 & 0     & 28.34 & 1.59  & C     & O \\
    F     & Mexico & Sweden & 81.94 & 0     & 5.71  & 93.90  & 81.94 & C     & AN \\ \hline
    G     & Belgium & England & 0     & 0     & 0     & 0     & 0     & S     & U \\
    G     & Panama & Tunisia & 0     & 0     & 0     & 0     & 0     & S     & U \\ \hline
    H     & Poland & Japan & 0     & 0     & 59.03 & 0     & 59.03 & AS    & DAS \\
    H     & Senegal & Colombia & 78.91 & 0     & 58.02 & 41.98 & 16.04 & C     & D \\ \bottomrule
    \end{tabularx}
\begin{tablenotes} \footnotesize
\item
Models: Det.\ stands for Deterministic, Prob.\ stands for Probabilistic.
\item
Deterministic match types: AS = asymmetric; C = competitive; COL = collusive; S = stakeless.
\item
Probabilistic match types: AN = antagonistic; D = defensive; DAS = defensive asymmetric; O = offensive; OAS = offensive asymmetric; U = unimportant.
\end{tablenotes}
\end{threeparttable}
}
\end{table}

\end{document}